\documentclass[aip,pop,preprint]{revtex4-2}
\usepackage{amsmath}
\usepackage{amssymb}
\usepackage{graphicx}
\usepackage{hyperref}
\graphicspath{{figures/}} 
\usepackage[usenames, dvipsnames]{color}

\newcommand{\Rm}{\mathrm{Rm}}

\newcommand{\phiphase}{\Delta_\phi}

\newcommand{\psiphase}{\Delta_\psi}

\begin{document}
	\author{A.E.~Fraser$^{1,2}$}
	\author{P.W.~Terry$^2$}
	\author{E.G.~Zweibel$^2$}
	\author{M.J.~Pueschel$^{3,4,5}$}
	\author{J.M.~Schroeder$^2$}
	\affiliation{$^1$University of California, Santa Cruz, Santa Cruz, California 95064, U.S.A.\\
	$^2$University of Wisconsin-Madison, Madison, Wisconsin 53706, U.S.A.\\
	$^3$Dutch Institute for Fundamental Energy Research, 5612 AJ Eindhoven, The Netherlands\\
	$^4$Eindhoven University of Technology, 5600 MB Eindhoven, The Netherlands\\
	$^5$Institute for Fusion Studies, University of Texas at Austin, Austin, Texas 78712, U.S.A.}
	\title{The impact of magnetic fields on momentum transport and saturation of shear-flow instability by stable modes}
	\begin{abstract}
		The Kelvin-Helmholtz (KH) instability of a shear layer with an initially-uniform magnetic field in the direction of flow is studied in the framework of 2D incompressible magnetohydrodynamics with finite resistivity and viscosity using direct numerical simulations. The shear layer evolves freely, with no external forcing, and thus broadens in time as turbulent stresses transport momentum across it. As with KH-unstable flows in hydrodynamics, the instability here features a conjugate stable mode for every unstable mode in the absence of dissipation. Stable modes are shown to transport momentum up its gradient, shrinking the layer width whenever they exceed unstable modes in amplitude. In simulations with weak magnetic fields, the linear instability is minimally affected by the magnetic field, but enhanced small-scale fluctuations relative to the hydrodynamic case are observed. These enhanced fluctuations coincide with increased energy dissipation and faster layer broadening, with these features more pronounced in simulations with stronger fields. These trends result from the magnetic field reducing the effects of stable modes relative to the transfer of energy to small scales. 
		As field strength increases, stable modes become less excited and thus transport less momentum against its gradient. Furthermore, the energy that would otherwise transfer back to the driving shear due to stable modes is instead allowed to cascade to small scales, where it is lost to dissipation. Approximations of the turbulent state in terms of a reduced set of modes are explored. While the Reynolds stress is well-described using just two modes per wavenumber at large scales, the Maxwell stress is not.
	\end{abstract}
	\maketitle

	\section{Introduction}
	\label{sec:intro}
	
	Shear layers are ubiquitous in space and astrophysical systems, 
	including Earth's magnetosphere\cite{Faganello}, relativistic jets \cite{Rieger}, and clouds passing by galactic and circumgalactic gas \cite{Kwak}. These flows exhibit extremely large Reynolds numbers, and thus are often susceptible to shear-flow instabilities that can give rise to turbulence. Among those instabilities is the Kelvin-Helmholtz (KH) instability, the canonical shear-flow instability \cite{Chandrasekhar,DrazinReid,Drazin}, which is triggered in unstable flow profiles by sufficiently strong flow shear (i.e.~it does not require other physical effects for it to be destabilized) and thus can exist in a wide range of systems. Despite the relative simplicity of the instability in idealized systems, the background magnetic fields present in many astrophysical systems can have significant and complex effects on its dynamics. If the field has a sufficiently strong component in the direction of flow, then KH is stabilized \cite{Chandrasekhar}. This stability threshold depends on fluid properties and the flow profile (see Ref.~\cite{Palotti} and references therein) but, in the perfectly conducting case, is roughly characterized by the Alfv\'{e}n velocity (in terms of the flow-aligned component of the field) exceeding the difference in flow velocity on either side of the layer. 
	For weaker magnetic fields, the instability remains but with a reduced growth rate. However, despite the magnetic field decreasing the growth rate relative to the hydrodynamic case, magnetohydrodynamic (MHD) simulations show even weak magnetic fields significantly enhance the generation of small-scale fluctuations and the rate at which momentum is transported across the shear layer, causing the layer to broaden at a faster rate \cite{Palotti,Mak}.
	
	This instability-driven turbulence generally transports momentum, as well as heat and particles, far faster than viscosity and molecular diffusion would alone. Thus, this turbulence can have important effects on systems where it is found. Indeed, transport driven by shear-flow turbulence is often necessary to explain observations \cite{BalbusHawley}, or similarly plays a key role in reduced dynamical models \cite{Paxton,Heger}. This motivates studies of shear-driven turbulence, particularly in pursuit of transport models that can be employed when considering systems too complex for direct numerical simulation (e.g.~in stellar evolution codes \cite{Paxton,Heger}). 
	With the significant impacts that background magnetic fields can have on turbulent transport in shear-driven turbulence, it is important that such transport models account for magnetic effects. 
	However, reduced models where transport is assumed to scale with the growth rate of the driving instability, and no details of the nonlinear saturation are included, are clearly inadequate in this case, as increasing magnetic field strength increases transport while decreasing the instability's growth rate. A primary goal of this work is to explore what details of the dynamics are responsible for this scaling, and thus might be necessary to include in reduced models to capture key trends accurately.
	
	In the context of turbulence driven by gyroradius-scale instabilities in fusion plasmas, novel reduced transport models \cite{Terry2018,Hegna}, as well as corrections to existing models \cite{WhelanPRL,WhelanPoP2019}, have been derived by accounting for the physical mechanisms that saturate the instability. Here, instability saturation refers to the arresting of the exponential growth of fluctuations that are seeded by initially-small perturbations. For example, in systems where perturbations grow by drawing energy from an unstable momentum, density, or temperature gradient, exponential growth might cease once perturbations have drawn so much energy from the driving gradient that it relaxes and is no longer unstable; this is sometimes referred to as quasilinear flattening. Particularly in systems with fixed background gradients, saturation might instead occur when the injection of energy by the instability is balanced by the nonlinear transfer of energy to small, dissipative scales. A third, distinct saturation mechanism involves the transfer of energy to large-scale, linearly stable (damped) modes \cite{Terry2006,Fraser2017}. These modes are eigenfunctions of the linearized governing equations, and they decay exponentially in the absence of nonlinear energy transfer from other modes. Signatures of their excitation due to nonlinear energy transfer have been measured in dipole-confined plasmas \cite{Qian}, and, in this paper, a previously-observed feature of hydrodynamic shear layers in laboratory experiments\cite{HoHuerre,Hurst2020} and simulations\cite{Takamure}, namely counter-gradient momentum transport, will be identified as a consequence of stable-mode excitation. 
	
	In the fusion context, analytical calculations have shown that stable-mode excitation is almost universally a significant contributor to saturation \cite{Terry2006,Makwana}. Subsequent direct numerical simulations have demonstrated that stable eigenmodes not only affect instability saturation, but also remain excited in the ensuing turbulence \cite{HatchLeft,TerryLeft}. Understanding the nonlinear interactions primarily responsible for energy transfer to stable modes \cite{Makwana2014} has enabled the development of a variety of reduced models that incorporate stable mode effects \cite{Terry2018,Hegna,WhelanPRL,WhelanPoP2019}. 
	In the case of shear-flow instabilities, the same analytical saturation calculations of Refs.~\cite{Terry2006,Makwana} were applied to a hydrodynamic, KH-unstable system in Ref.~\cite{Fraser2017}, showing that stable modes are important in saturating the KH instability, and that when they are excited they can significantly affect turbulent momentum transport (see also Ref.~\cite{FraserThesis}, where the methods were extended to systems with eigenmodes that cannot be derived in closed form, like the system considered here, and thus the calculation must be done numerically). In Ref.~\cite{Fraser2018}, gyrokinetic simulations of an unstable shear flow revealed that stable modes are excited to significant amplitudes in shear-driven turbulence except when heavily damped by a radiative damping term (also called drag or friction in other contexts). There, an external forcing term partially maintained the unstable flow profile against quasilinear flattening, permitting a quasi-stationary state of driven turbulence. 
	The simulations had only two spatial dimensions, and the dynamics were essentially hydrodynamic, with the forcing term \cite{PueschelKrook} reminiscent of the forcing considered in the well-studied Kolmogorov flow problem \cite{Platt,Musacchio,Lucas}. In parameter regimes where stable modes were significantly excited, the relevant flow fluctuations could be approximated well by linear combinations of stable and unstable modes alone (neglecting the continuum of marginally stable modes \cite{Case}). A scaling model for a quantity directly related to the Reynolds stress as a function of the forcing was derived in terms of the stable and unstable mode amplitudes and shown by comparison with simulations to be very accurate.
	
	The present work explores the role of stable modes in shear-flow instability saturation and turbulent momentum transport for a system that differs from Ref.~\cite{Fraser2018} in two key regards. First, no forcing terms are included. Thus, no quasi-stationary state is formed, quasilinear flattening is permitted, and the effects of layer broadening on saturation and the ensuing turbulence are investigated. Second, an initially-uniform magnetic field in the direction of flow is included. The present study focuses on the weak-field regime, where the growth rate of the instability is only slightly reduced compared to the hydrodynamic case. The system is studied in the MHD framework via direct numerical simulations using the code Dedalus \cite{Dedalus,Lecoanet}. The simulations are of a two-dimensional (2D), incompressible fluid with finite viscosity and resistivity included explicitly. 
	
	In other systems of instability-driven turbulence, stable modes are known to remove energy from fluctuations that would otherwise cascade to small scales \cite{Hatch2013,Makwana2014}. Here, the enhancement of turbulent momentum transport and small-scale fluctuations with increasing magnetic field strength is examined in the context of stable modes to identify whether the enhanced small-scale fluctuations are due to a reduction in stable-mode activity. At large scales, the linearized, dissipationless system features a pair of unstable and stable modes at each horizontal wavenumber \cite{Chandrasekhar}. Unstable modes gain energy from the shear flow, and their associated Reynolds and Maxwell stresses transport momentum down its gradient and broaden the layer \cite{Drazin}. Stable modes return energy to the background, and their stresses transport momentum against its gradient and tend to shrink the layer. Transient instances of counter-gradient momentum transport, resembling those seen in experiments \cite{HoHuerre,Hurst2020}, are observed here and shown to occur whenever stable modes exceed unstable modes in amplitude. 
	As field strength is increased between simulations, this counter-gradient momentum transport becomes weaker and eventually ceases partly because stable modes are less excited with stronger magnetic fields. With less stable mode activity, the energy that would otherwise be returned to the background flow instead cascades to small scales, producing more small-scale fluctuations and a significant increase in energy dissipation relative to the hydrodynamic case. The small-scale fluctuations also produce a down-gradient Maxwell stress that becomes significant for strong initial fields or sufficiently low resistivity. The enhanced layer broadening is thus a combination of reduced counter-gradient momentum transport by stable modes and enhanced down-gradient transport by small-scale magnetic fluctuations.
	
	This paper is organized as follows. The system is described in Sec.~\ref{sec:setup}, including the equilibrium, governing equations, and the numerical implementation. The dissipationless linear modes are discussed in Sec.~\ref{sec:eigenmodes}. Nonlinear simulations are presented in Sec.~\ref{sec:nonlinear}, beginning with an overview of the nonlinear evolution of the system, followed by discussions of the effects of layer broadening in Sec.~\ref{nonlinear:subsec:broadening}, stable-mode excitation and momentum transport in Sec.~\ref{nonlinear:subsec:ModeExcitationTransport}, and small-scale fluctuations and dissipation in Sec.~\ref{nonlinear:subsec:SmallScales}. Conclusions are presented in Sec.~\ref{sec:conclusions}.

	\section{System setup}
	\label{sec:setup}
	
	\subsection{Equilibrium, governing equations}
	\label{setup:subsec:equilibrium}
	
	We study the evolution of a two-dimensional free shear layer in an incompressible fluid in MHD with finite viscosity and resistivity, with an initially-uniform magnetic field in the direction of the flow. Specifically, we consider an initial flow in the horizontal direction $\hat{\mathbf{x}}$ that varies in the vertical direction $\hat{\mathbf{z}}$, i.e., $\bar{\mathbf{V}}_0 = \bar{U}(\bar{z}) \hat{\mathbf{x}}$, where $\bar{U}(\bar{z}) = \bar{U}_0\tanh(\bar{z}/\bar{d})$ is the initial flow profile, $\bar{U}_0$ is the flow speed away from the layer, and $\bar{d}$ is the layer half-width, with an initial, uniform magnetic field $\bar{\mathbf{B}}_0 = \bar{B}_0 \hat{\mathbf{x}}$. Here, an overbar denotes a dimensional quantity. Henceforth, we non-dimensionalize all speeds, distances, and fields according to $U = \bar{U}/\bar{U}_0$, $(x,z) = (\bar{x}/\bar{d}, \bar{z}/\bar{d})$, and $B = \bar{B}/\bar{B}_0$, respectively, such that $\mathbf{V}_0 = \tanh(z) \hat{\mathbf{x}}$ and $\mathbf{B}_0 = \hat{\mathbf{x}}$. All other physical quantities will be non-dimensionalized in terms of $\bar{U}_0$, $\bar{d}$, $\bar{B}_0$, and combinations thereof. 
	
	We describe the flow velocity and magnetic field in terms of a streamfunction $\phi$ and a flux function $\psi$, so that $\mathbf{v} = \hat{\mathbf{y}} \times \nabla \phi$ and $\mathbf{B} = \hat{\mathbf{y}} \times \nabla \psi$. Under our chosen non-dimensionalization, we may write the governing equations as \cite{Biskamp}
	\begin{equation}\label{eq:NLVorticity}
	\frac{\partial}{\partial t} \nabla^2 \phi + \left\{ \nabla^2 \phi, \phi \right\} = \frac{1}{M_\mathrm{A}^2}\left\{ \nabla^2 \psi, \psi \right\} + \frac{1}{\mathrm{Re}}\nabla^4 \phi
	\end{equation}
	and
	\begin{equation}\label{eq:NLFlux}
	\frac{\partial}{\partial t}\psi = \left\{ \phi, \psi \right\} + \frac{1}{\mathrm{Rm}}\nabla^2 \psi.
	\end{equation}
	Here, $M_\mathrm{A}$ is the Alfv\'{e}n Mach number, or the ratio of the equilibrium flow speed to the Alfv\'{e}n speed, and scales like $M_\mathrm{A} \propto \bar{U}_0/\bar{B}_0$; the Reynolds number $\mathrm{Re}$ and magnetic Reynolds number $\mathrm{Rm}$ are defined as $\mathrm{Re} = \bar{U}_0 \bar{d}/\bar{\nu}$ and $\Rm = \bar{U}_0 \bar{d}/\bar{\mu}$, respectively, where $\bar{\nu}$ is the kinematic viscosity and $\bar{\mu}$ is resistivity; 
	and $\left\{ f, g \right\} \equiv \partial_x f \partial_z g - \partial_x g \partial_z f$. 
	Equation \eqref{eq:NLVorticity} describes the evolution of the vorticity $\nabla \times \mathbf{v} = \nabla^2 \phi \hat{\mathbf{y}}$. The second term on the left-hand side is the vorticity advection term, the first term on the right-hand side is the curl of the Lorentz force, and the second term on the right-hand side is standard viscous dissipation. The terms on the right-hand side of Eq.~\eqref{eq:NLFlux} correspond to flux advection and resistive diffusion. 
	This system, with the above equilibrium, is known to be linearly unstable for $M_\mathrm{A}$ above a critical threshold that lies between $1$ and $2$ \cite{Palotti}.
	
	\subsection{Perturbation equations}
	\label{setup:subsec:perturbation}
	As will be described in Sec.~\ref{setup:subsec:implementation}, we solve Eqs.~\eqref{eq:NLVorticity} and \eqref{eq:NLFlux} numerically using the initial value problem capabilities in the Dedalus code \cite{Dedalus}. Additionally, we use Dedalus' eigenvalue problem capabilities to calculate the complex frequencies (eigenvalues) and eigenmodes of these equations linearized about an unstable equilibrium. 
	Solving the linearized system as an eigenvalue problem allows the full set of eigenmodes at each $k_x$, including stable modes, to be calculated, whereas initial value calculations yield only the most unstable mode at each $k_x$. This is necessary to track the amplitudes of these modes in the ensuing turbulence in solutions of Eqs.~\eqref{eq:NLVorticity} and \eqref{eq:NLFlux}, which then informs how much energy they remove from fluctuations \cite{Makwana2014}. 
	As with previous studies of stable modes in shear-flow turbulence \cite{Fraser2017, Fraser2018}, we are specifically interested in the dissipationless modes of this system, so eigenmodes are calculated with viscosity and resistivity neglected. While eigenmodes could be calculated with dissipation included, such modes would mix together the physical effects of conservative energy transfer between the shear flow and fluctuations -- a primary focus of this paper -- and non-conservative dissipation at large scales. The stable modes of the dissipative system owe their stability to a combination of these two effects, and thus these modes do not lend themselves as conveniently to calculations involving the conservative effect alone. Furthermore, the dissipationless modes have previously been shown to still be relevant in the full, dissipative system \cite{Fraser2018,HatchPseudo}.
	
	To derive linearized equations, we separate the system into a horizontal, uniform (in $x$) background flow $U(z)$ and field $B_x(z)$, and perturbations $\tilde{\phi}$ and $\tilde{\psi}$.
	This allows Eqs.~\eqref{eq:NLVorticity} and \eqref{eq:NLFlux} to be similarly separated into equations describing the background, and the following equations for the perturbations: 
	\begin{equation}\label{eq:NLphipert}
	\begin{split}
	\frac{\partial}{\partial t} \nabla^2 \tilde{\phi} = &-U \frac{\partial}{\partial x} \nabla^2 \tilde{\phi} + U'' \frac{\partial}{\partial x} \tilde{\phi} + \frac{1}{M_\mathrm{A}^2}\left( B_x \frac{\partial}{\partial x} \nabla^2 \tilde{\psi} - B_x'' \frac{\partial}{\partial x} \tilde{\psi} \right)\\ &- \left\{ \nabla^2 \tilde{\phi}, \tilde{\phi} \right\} + \frac{1}{M_\mathrm{A}^2} \left\{ \nabla^2 \tilde{\psi}, \tilde{\psi} \right\}
	\end{split}
	\end{equation}
	and
	\begin{equation}\label{eq:NLpsipert}
	\frac{\partial}{\partial t} \tilde{\psi} = - U \frac{\partial}{\partial x} \tilde{\psi} + B_x \frac{\partial}{\partial x} \tilde{\phi} + \left\{ \tilde{\phi}, \tilde{\psi} \right\},
	\end{equation}
	where primes denote derivatives with respect to $z$, and we have neglected viscosity and resistivity. 
	Equations \eqref{eq:NLphipert} and \eqref{eq:NLpsipert} are the MHD equivalent of Eq.~(1) in Ref.~\cite{Fraser2017}, and describe how fluctuations interact linearly with the background flow and field, and nonlinearly with one another. When the fluctuations are small enough that the nonlinearities can be neglected, Fourier transforming Eqs.~\eqref{eq:NLphipert} and \eqref{eq:NLpsipert} in $x$ and assuming solutions vary in time as $\exp[i \omega(k_x) t]$ yields
	\begin{equation}\label{eq:linphieq}
	\begin{split}
	\omega \left( \frac{d^2}{dz^2} - k_x^2 \right) \hat{\phi} = &-k_x U \left( \frac{d^2}{dz^2} - k_x^2 \right) \hat{\phi} + k_x U'' \hat{\phi}\\ &+ \frac{1}{M_\mathrm{A}^2}\left[ k_x B_x \left( \frac{d^2}{dz^2} - k_x^2 \right) \hat{\psi} - k_x B_x'' \hat{\psi} \right]
	\end{split}
	\end{equation}
	and
	\begin{equation}\label{eq:linpsieq}
	\omega \hat{\psi} = -k_x U \hat{\psi} + k_x B_x \hat{\phi},
	\end{equation}
	where $\hat{\phi}$ and $\hat{\psi}$ are the Fourier transforms (in $x$) of $\tilde{\phi}$ and $\tilde{\psi}$. 
	Thus, at every $k_x$, we have a separate system of linear, ordinary differential equations in $z$. For a given $k_x$ and $M_\mathrm{A}$, $U(z)$ and $B_x(z)$, and appropriate choice of boundary conditions, this system forms a generalized eigenvalue problem that can be solved to obtain a spectrum of eigenvalues $\omega_j$ and eigenmodes $\vec{f}_j \equiv (\phi_j(z), \psi_j(z))$, where $j=1,2,\dots$ enumerates the different solutions at each $k_x$. While the equilibrium considered in this paper is specifically $U = \tanh(z)$ and $B_x = 1$, the more general equations are presented here because eigenmodes corresponding to other $U(z)$ and $B_x(z)$ will be considered as well in this paper. 
	
	\subsection{Numerical implementation}
	\label{setup:subsec:implementation}
	Dedalus is a pseudo-spectral code with a variety of spectral bases available. We employ Fourier modes $\exp[i k_x x]$ in the $x$ direction and Chebyshev polynomials $T_n(z)$ in $z$. Our simulation domain size is $L_x \times L_z = 10 \pi \times 10 \pi$, thus the minimum horizontal wavenumber is $k_x=0.2$, with periodic boundaries at $x = \pm L_x/2$ and perfectly conducting, no-slip, co-moving (with the equilibrium flow $\mathbf{V}_0$) walls at $z = \pm L_z/2$. The simulations presented here use a resolution of $N_x \times N_z = 512 \times 2048$, with convergence tests performed at the highest values of $\mathrm{Rm}$ by ensuring that changes in spectral energy density and dissipation with resolution are minimal. 
	For many of the physical parameters studied here, this $z$-resolution is higher than necessary for well-resolved simulations. However, some of our eigenmode-based post-processing analyses benefit from increased $N_z$, as it allows continuum modes \cite{Case} to be better-resolved. 
	
	Previous work has shown that the nonlinear development of KH-unstable flows depends sensitively on the choice of the initial perturbations that seed the instability \cite{Dong2019}. 
	In studying free shear layers, a common choice of initial condition is a perturbation in one or more velocity fields that is sinusoidal in $x$, with a wavelength that matches the box size or the fastest-growing linear mode, and Gaussian in $z$ centered about the shear layer \cite{Lecoanet, Palotti}, with lower-amplitude noise sometimes added to other horizontal wavenumbers \cite{Mak}. 
	Here, we perturb both $\phi$ and $\psi$ at every nonzero $k_x$ with Gaussians in $z$ that have randomly-assigned, $k_x$-dependent complex phases and amplitudes that decrease with $k_x$ as a power law. Thus, at $t=0$ the streamfunction and flux function are 
	\begin{equation}\label{eq:phiIC}
	\phi(x, z) = \sum_{k_x} \hat{\phi}(k_x, z) = \hat{\phi}(0, z) + A_\phi \sum_{k_x>0} k_x^{a} e^{i \phiphase (k_x) - z^2/\sigma^2},
	\end{equation}
	and
	\begin{equation}\label{eq:psiIC}
	\psi(x, z) = \sum_{k_x} \hat{\psi}(k_x, z) = \hat{\psi}(0, z) + A_\psi \sum_{k_x>0} k_x^{a} e^{i \psiphase (k_x) - z^2/\sigma^2},
	\end{equation}
	where the $k_x=0$ components are the unperturbed equilibrium profiles, $A_\phi$ and $A_\psi$ set overall amplitudes for the perturbations, $a$ sets the steepness of the energy spectra of the perturbations, $\sigma$ sets the width of the Gaussian in $z$, and at every nonzero wavenumber, $\phiphase(k_x)$ and $\psiphase(k_x)$ are uniformly-distributed pseudo-random numbers in $[0, 2\pi)$. For the results presented here, we use $\sigma = 2$, 
	$a = -1$, and $A_\phi = A_\psi = 5 \times 10^{-4}$, which allows for a clearly-defined regime of linear growth before nonlinear interactions become important. For $M_\mathrm{A} = 5$, setting $A_\psi = 0$ did not noticeably change how the instability saturated. This is likely a result of the flow-dominated nature of the instability at these values of $M_\mathrm{A}$ (as will be shown in Sec.~\ref{sec:eigenmodes}) and the well-defined linear growth regime permitted by our small value of $A_\phi$.
	
	Previous work \cite{Dong2019} has shown that even when only two wavenumbers are perturbed, the details of the nonlinear stage after the instability saturates are sensitive to the complex phase differences and relative amplitudes between different $k_x$, the overall amplitude of the perturbation, and the structure in $z$ of the perturbations. In this work, we are interested in studying details of the saturated state as $M_\mathrm{A}$ and $\mathrm{Rm}$ are varied. In an effort to ensure that our observed trends are not a unique feature of a particular choice of initial conditions, we perform multiple simulations at each $M_\mathrm{A}$ and $\mathrm{Rm}$, with different ensembles of $\phiphase$ and $\psiphase$. In practice, this is done by selecting different seeds for our pseudo-random number generator (we use \texttt{numpy.random.RandomState} \cite{numpy,scipy} to ensure consistency across different computers) 
	and using the same seeds for different $M_\mathrm{A}$ and $\Rm$ so that $M_\mathrm{A}$ and $\mathrm{Rm}$ can be varied independently with $\phiphase$ and $\psiphase$ held fixed. For each value of $M_\mathrm{A}$ and $\Rm$ presented here, at least five different sets of initial conditions were simulated. While the majority of this paper presents results from only one set of initial conditions, the trends we present were robust and broadly representative of the range of initial conditions we sampled.

	\section{Eigenmodes and eigenvalues for $U = \tanh (z)$ and $B_x=1$}
	\label{sec:eigenmodes}
	
	\begin{figure}
		\includegraphics[width=0.75\textwidth]{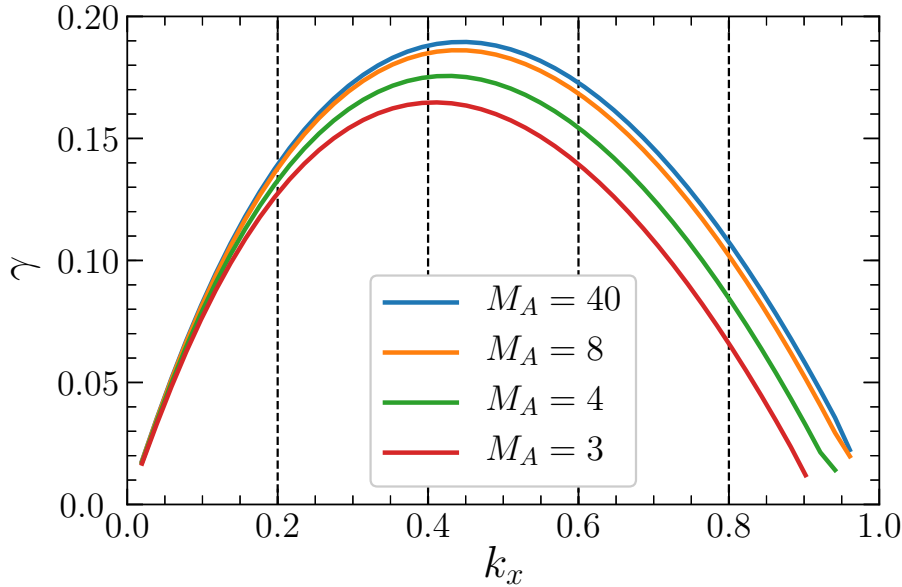}
		\caption[Dispersion relation, MHD shear layer]{Growth rate $\gamma$ for the fastest-growing mode at each $k_x$. Each curve corresponds to a different Alfv\'{e}n Mach number $M_\mathrm{A}$. While stronger magnetic fields (lower $M_\mathrm{A}$) provide a stabilizing influence, $\gamma$ varies little except when $M_\mathrm{A} \lesssim 4$. Horizontal dashed lines indicate the $k_x$ present in our nonlinear simulations.}\label{fig:dispersion}
	\end{figure}
	
	For $U = \tanh(z)$ and $B_x=1$, unstable modes, solutions to Eqs.~\eqref{eq:linphieq} and \eqref{eq:linpsieq} with positive growth rates $\gamma_j = - \mathrm{Im}[\omega_j]$, are observed as expected for wavenumbers in the range $0 < k_x < 1$ as long as $M_\mathrm{A}$ is above a critical threshold between $1$ and $2$ (the precise value depends on fluid properties and the flow profile, see Ref.~\cite{Palotti} and references therein). The growth rate of the fastest-growing mode for this system is plotted against $k_x$ for a variety of $M_\mathrm{A}$ in Fig.~\ref{fig:dispersion}. 
	Magnetic tension provides a stabilizing influence that suppresses instability for $M_\mathrm{A}$ below the critical threshold, and significantly reduces the growth rate for $M_\mathrm{A}$ slightly above the threshold, but only marginally affects the growth rate for $M_\mathrm{A} \gtrsim 8$.
	
\begin{figure}
		\includegraphics[width=16cm]{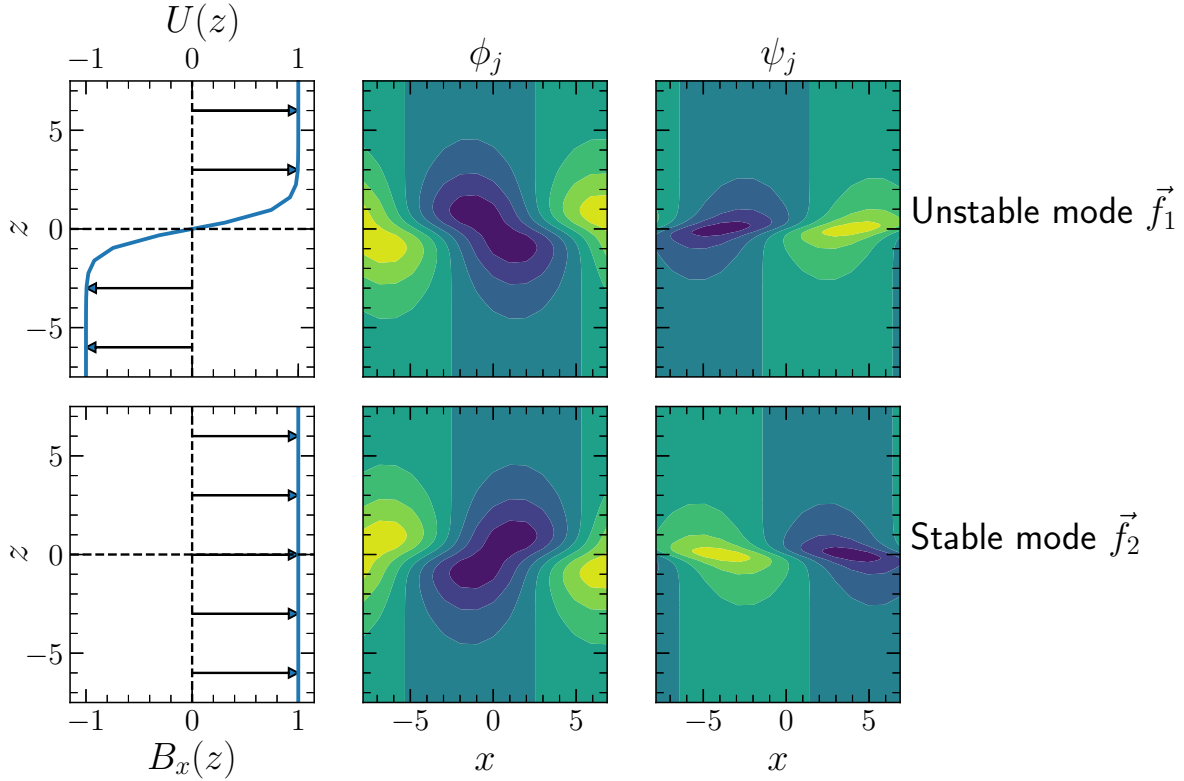}
		\caption[MHD Eigenmodes]{The initial $U = \tanh (z)$ unstable equilibrium flow (top left) and uniform field (bottom left) are shown alongside contours (with arbitrary units) of the streamfunction (center column) and flux function (right column) for the unstable (top) and stable (bottom) modes at $k_x=0.4$ for $M_\mathrm{A}=40$.} 
		\label{fig:unstable_cont}
	\end{figure}

	Taking the complex conjugate of Eqs.~\eqref{eq:linphieq} and \eqref{eq:linpsieq} shows that, as in the hydrodynamic \cite{DrazinReid} and gyrokinetic cases \cite{Fraser2018}, for every eigenvalue $\omega_j$ and eigenmode $(\phi_j, \psi_j)$ that is a solution of Eqs.~\eqref{eq:linphieq} and \eqref{eq:linpsieq}, the complex conjugate $\omega_j^*$ and $(\phi_j^*, \psi_j^*)$ is a solution as well. Following Refs.~\cite{Fraser2017,Fraser2018}, when describing the eigenmodes of this system, we label the most unstable mode at each $k_x$ as $j=1$ and its conjugate stable mode as $j=2$. Real-space contours corresponding to $\phi_j$ and $\psi_j$ for $j=1$ and $j=2$ at $M_\mathrm{A} = 40, k_x=0.4$ are shown in Fig.~\ref{fig:unstable_cont}. 
	While the flow component of the mode can be roughly described as a superposition of two waves of vorticity localized about the edges of the layer (a wealth of literature exists on this subject \cite{Baines,Heifetz2015,Heifetz2019,Carpenter}), the current density of the mode is more localized about the center of the layer. 
	
	The source of free energy that drives the exponential growth of the unstable mode is the equilibrium flow $U(z)$. In terms of Eqs.~\eqref{eq:NLphipert} and \eqref{eq:NLpsipert}, the growth of the mode is due to the (dissipationless) linear terms on the on right-hand side. These terms were derived from the energy-conserving nonlinearities in Eqs.~\eqref{eq:NLVorticity} and \eqref{eq:NLFlux} by separating interactions involving $U(z)$ and $B_x(z)$ from nonlinear interactions between fluctuations. If $U(z)$ and $B_x(z)$ are identified as the horizontally-averaged flow and field and held fixed in time, so that only $k_x \neq 0$ perturbations are allowed to evolve, then this exponential growth does not conserve energy, because the energy injected into $\vec{f}_1$ is not self-consistently removed from the mean flow. This is the case in Ref.~\cite{Fraser2017}, as well as previous studies of stable modes in plasma turbulence driven by instabilities aside from KH due to the fixed background gradients that were considered (e.g.~Refs.~\cite{Terry2006,Makwana,HatchLeft,TerryLeft,Makwana2014}). However, in direct numerical simulations of Eqs.~\eqref{eq:NLVorticity} and \eqref{eq:NLFlux} (or in other systems where driving gradients are not held fixed), energy is conservatively transferred from the equilibrium to growing perturbations by the nonlinearities. These considerations are critical for the relationship between the horizontally averaged flow, eigenmodes, and Reynolds stress in counter-gradient transport events described in Sec.~\ref{sec:nonlinear}. Viewed in terms of a separation between the mean and $k_x \neq 0$ fluctuations, the removal of energy from $U(z)$ occurs via the $xz$ components of the Reynolds and/or Maxwell stress tensors, which we denote as
	\begin{equation}\label{eq:tau_u}
	\tau_u \equiv - \left\langle \frac{\partial}{\partial x} \tilde{\phi} \frac{\partial}{\partial z} \tilde{\phi} \right\rangle_x
	\end{equation}
	and 
	\begin{equation}\label{eq:tau_b}
	\tau_b \equiv \frac{1}{M_\mathrm{A}^2}\left\langle \frac{\partial}{\partial x} \tilde{\psi} \frac{\partial}{\partial z} \tilde{\psi} \right\rangle_x,
	\end{equation}
	respectively, where $\langle \cdot \rangle_x$ indicates an average in $x$. 
	These stresses transport horizontal momentum along the vertical axis and evolve the mean flow according to (neglecting viscosity)
	\begin{equation}\label{eq:meanflow_evolution}
	\frac{\partial}{\partial t} \langle U \rangle_x = \frac{\partial}{\partial z} \left(\tau_u + \tau_b\right),
	\end{equation}
	with a transport of momentum down the gradient lowering the kinetic energy of the mean flow.
	
	\begin{figure}
		\includegraphics[width=16cm]{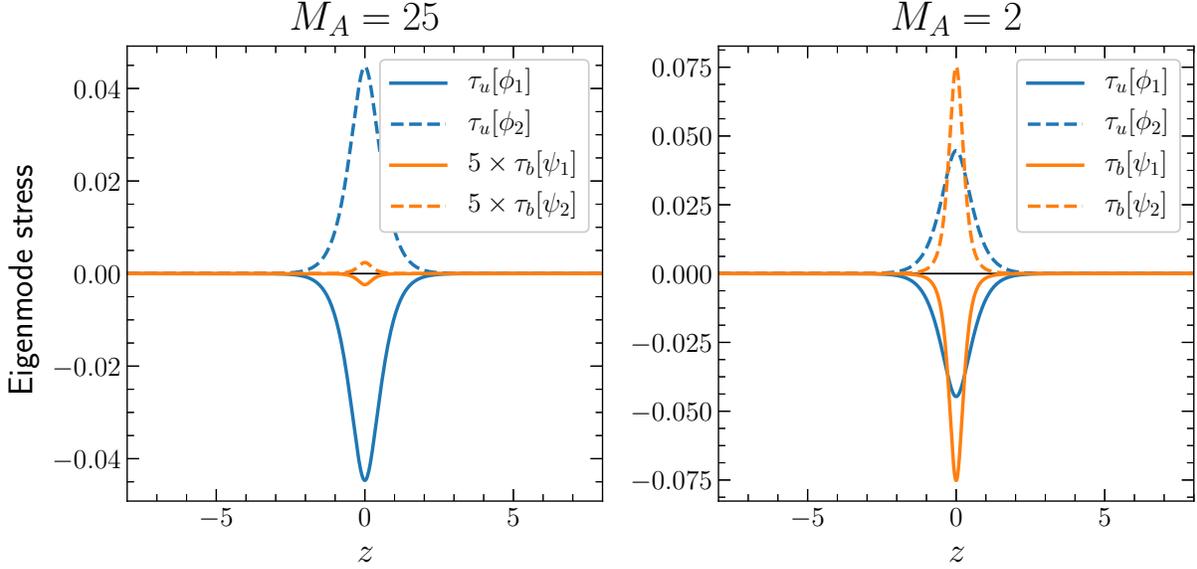}
		\caption[Eigenmode Reynolds and Maxwell stresses]{$xz$ components of the Reynolds stress $\tau_u$ (blue) and Maxwell stress $\tau_b$ (orange) for the unstable mode $\vec{f}_1$ (solid lines) and stable mode $\vec{f}_2$ (dashed lines) at $k_x=0.4$ for $M_\mathrm{A} = 25$ (left) and $M_\mathrm{A} = 2$ (right). For $M_\mathrm{A} = 25$, $\tau_b$ is rescaled by a factor of $5$ to improve visibility. Modes are normalized to have unit total energy.}
		\label{fig:eigenmode_stresses}
	\end{figure}
	
	As with the exponential growth of the unstable mode $\vec{f}_1$, the exponential decay of the conjugate stable mode $\vec{f}_2$ does not conserve energy if the background flow and field are held fixed. Thus, in Ref.~\cite{Fraser2017} and previous studies of stable modes in instability-driven turbulence, stable modes necessarily present a nonconservative energy sink. However, they do conserve energy when directly simulating Eqs.~\eqref{eq:NLVorticity} and \eqref{eq:NLFlux}, with energy injection into the mean provided by the same stresses, in addition to a minimal amount of energy that is transferred into the mean field. This is illustrated in Fig.~\ref{fig:eigenmode_stresses}, where the Reynolds and Maxwell stresses are shown for $\vec{f}_1$ and $\vec{f}_2$ at $k_x=0.4$ for both $M_\mathrm{A} = 25$ and $M_\mathrm{A} = 2$. For unstable modes, both $\tau_u$ and $\tau_b$ transport momentum down the gradient, so that they both contribute to a transfer of energy from $U(z)$ to $\vec{f}_1$. Likewise, for stable modes, both stresses yield counter-gradient momentum transport, transferring energy from $\vec{f}_2$ to $U(z)$. The transport of the two modes is symmetric in the sense that $\tau_u [\phi_2] = - \tau_u [\phi_1]$ and $\tau_b [\psi_2] = - \tau_b [\psi_1]$. As $M_\mathrm{A}$ is decreased, corresponding to a stronger equilibrium field, the relative amplitudes of $\tau_u$ and $\tau_b$ change, with $|\tau_b|$ exceeding $|\tau_u|$ for only the strongest equilibrium fields, starting around $M_\mathrm{A} \approx 2.5$.

	\section{Nonlinear evolution}\label{sec:nonlinear}
	\begin{figure}
		\includegraphics[width=\textwidth]{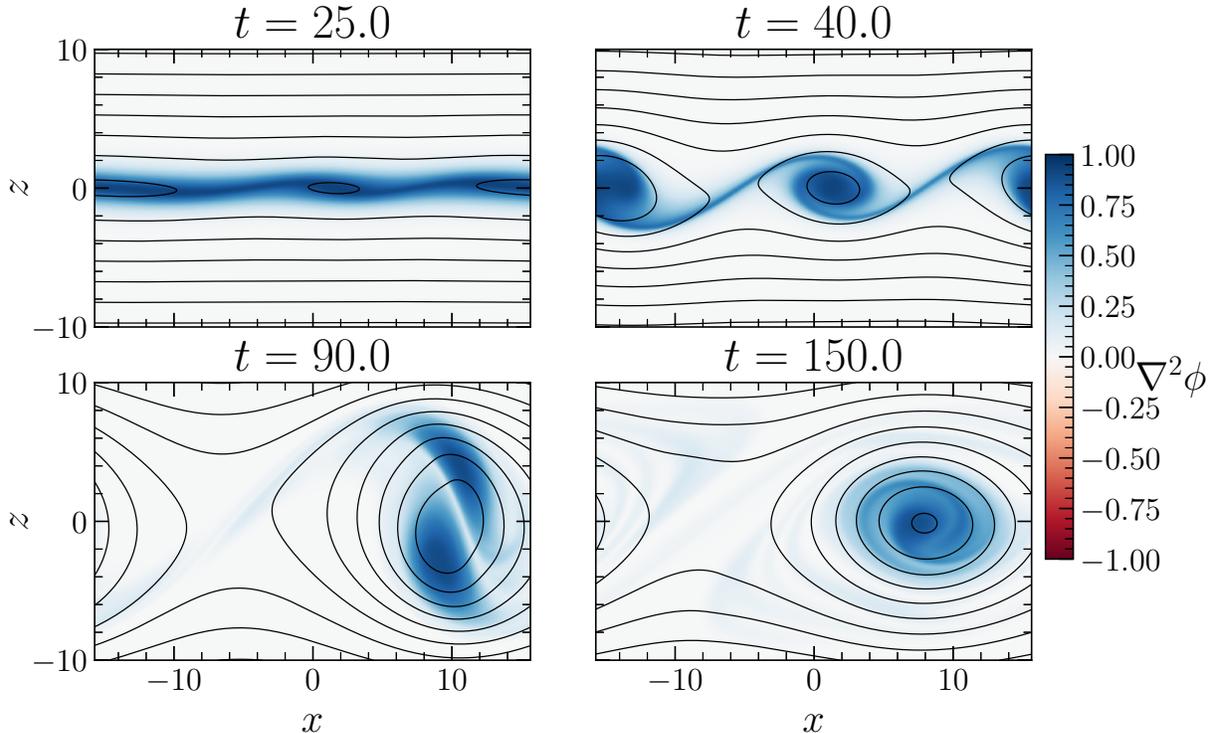}
		\caption{Snapshots of the vorticity and streamlines near the shear layer at four different times for a hydrodynamic simulation. Color shows vorticity $\nabla^2 \phi$, black lines show contours of the streamfunction $\phi$, representing streamlines of the flow. 
		Colorbar rescaled with $\nabla^2 \phi = 1$ as the maximum and $-1$ as the minimum with white as $0$ to demonstrate that, here, all of the vorticity is into the page, consistent with the conservative, advection-diffusion nature of the vorticity equation.}\label{fig:hydroNLcontours}
	\end{figure}
	
\begin{figure}
		\includegraphics[width=\textwidth]{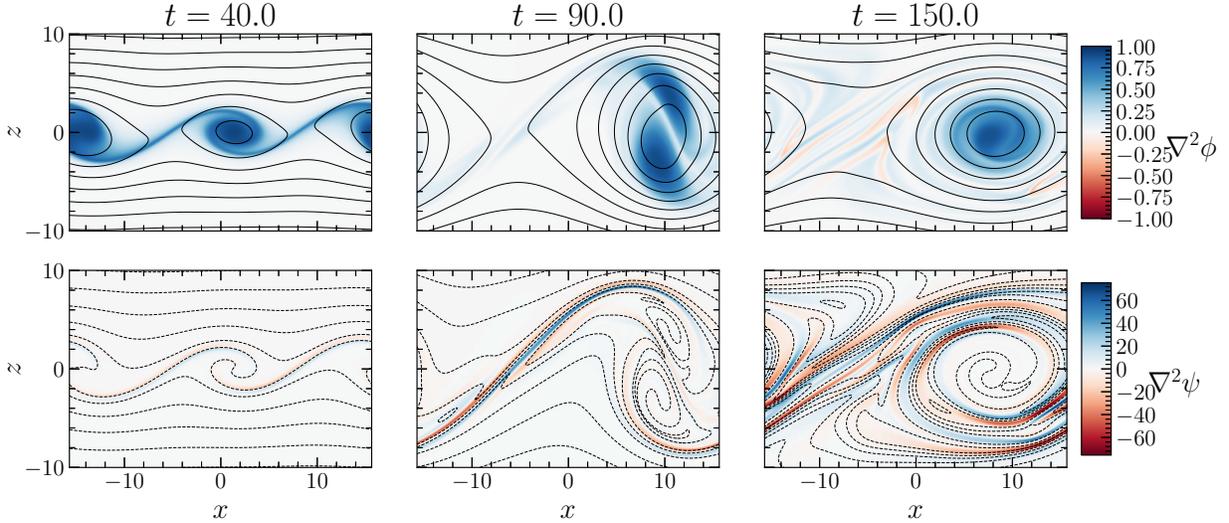}
		\caption[Flow and field snapshots, MHD]{Snapshots of the flow (top) and field (bottom) near the shear layer at three times in a simulation with $M_\mathrm{A} = 60$ and $\Rm = 250$. Black lines are contours of $\phi$ (top) and $\psi$ (bottom), representing streamlines and field lines, and color represents vorticity (top) and current density (bottom) into the page. Current sheets form early in the simulation along the braids between vortices, as well as at the edges of the vortex at later times. Even at this high $M_\mathrm{A}$ and low $\Rm$, the Lorentz force breaks vorticity conservation and introduces some negative vorticity late in the simulation, presenting a departure from the hydrodynamic evolution.}\label{fig:contours_26_1}
	\end{figure}

	Consistent with previous work \cite{Palotti,Mak}, the addition of even a weak magnetic field causes significant changes to the nonlinear evolution of this system despite only slight changes to the linear instability. This is readily seen by inspecting snapshots of the flow. Figure \ref{fig:hydroNLcontours} shows vorticity and streamlines for a hydrodynamic simulations at four different times, roughly corresponding to the linear growth phase, precursor vortex formation, vortex merging, and deep in the nonlinear regime. For comparison, Fig.~\ref{fig:contours_26_1} shows vorticity and streamlines, as well as current density and field lines, for an MHD simulation with the same initial conditions. This simulation has an initially-weak magnetic field, with $M_\mathrm{A} = 60$ and $\Rm = 250$. In the hydrodynamic case, the vorticity equation becomes an advection-diffusion equation, so no negative vorticity is produced aside from boundary layer effects. Vorticity conservation is broken by the Lorentz force in MHD, as can be seen by the regions of negative vorticity that form late in the simulation.
	
	Magnetic fields embedded in high-strain-rate flows, such as the narrow vortex sheets, known as braids, connecting the coherent vortices in this system, are known to be amplified by the strain provided $\Rm \gg 1$ \cite{ZweibelStagnation,Zeldovich}. The flow strain pushes neighboring field lines together, forming a current sheet along the braid as seen in Fig.~\ref{fig:contours_26_1}. This figure also demonstrates the familiar amplification of magnetic fields by coherent vortices \cite{Mak}, with current sheets forming at the edge of the post-merger vortex as it wraps up the magnetic field, with the field lines in the interior of the vortex reconnecting and drifting outwards via flux expulsion \cite{Weiss}. The field amplification provided by both high-strain-rate flow and coherent vortices increases with $\Rm$ \cite{ZweibelStagnation,Mak}. Thus, when studying trends with magnetic field strength in this system, not only does field strength vary with $M_\mathrm{A}$, but also with $\Rm$. Decreasing $M_\mathrm{A}$ corresponds to a stronger initial field, while increasing $\Rm$ allows the field to become more amplified as time goes on.

	\begin{figure}
		\includegraphics[width=1.0\textwidth]{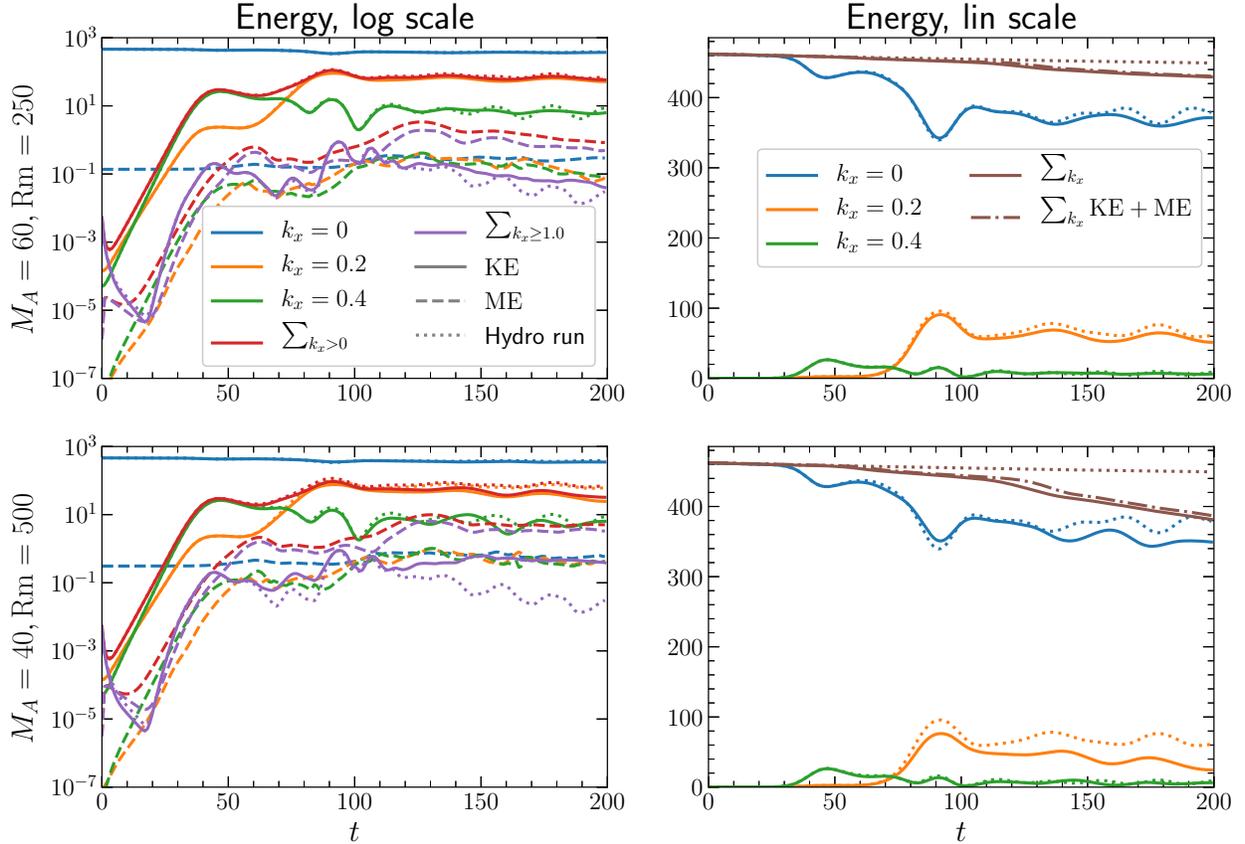}
		\caption[Energy timetraces, MHD]{Energy components versus time for the simulation shown in Fig.~\ref{fig:contours_26_1} with $M_\mathrm{A} = 60$ and $\Rm = 250$ (top row), and a similar simulation with $M_\mathrm{A} = 40$ and $\Rm = 500$ (bottom row). The left column plots energy on a log scale and the right on a linear scale. Solid, dashed, and dot-dashed curves correspond to kinetic, magnetic, and total energy, respectively. Dotted curves correspond to the hydrodynamic case (identical between top and bottom rows). Colors indicate different $k_x$ contributions. The initial saturation and merger stages show no obvious changes from the hydrodynamic case, but local minima in $\mathrm{KE}_{k_x=0}$ (blue) become less pronounced with sronger fields. Small-scale fluctuations become enhanced with increased field strength.}\label{fig:energies_26_9_1}
	\end{figure}
	
	Consistent with Ref.~\cite{Dong2019}, the precursor vortex formation time and the vortex merging time correspond to the initial saturation of $k_x=0.4$ and $k_x=0.2$, respectively. This is demonstrated in Fig.~\ref{fig:energies_26_9_1}, where various components of energy in the system are plotted over time, with solid and dashed lines corresponding to kinetic and magnetic energy for the MHD case, dotted lines corresponding to the hydrodynamic case, and different colors corresponding to different components of the 1D spectral energy density, defined so that $\mathrm{KE} = \sum_{k_x} \mathrm{KE}_{k_x}$ and $\mathrm{ME} = \sum_{k_x} \mathrm{ME}_{k_x}$. The precursor vortices form roughly when $\mathrm{KE}_{k_x=0.4}$ reaches its first maximum, and they merge roughly when $\mathrm{KE}_{k_x=0.2}$ reaches its first maximum. Consistent with Ref.~\cite{Dong2019}, we found the saturation time for $k_x=0.4$ to be independent of the complex phase $\phiphase(k_x)$ of the initial flow perturbation, while the saturation time for $k_x=0.2$ does depend on $\phiphase(k_x)$. Varying $\psiphase$ had no significant effect on the simulations. The simulations shown here correspond to a choice of $\phiphase(k_x)$ with a relatively long merging time.
	
	\subsection{Layer broadening}\label{nonlinear:subsec:broadening}
	
\begin{figure}
		\begin{center}
			\includegraphics[width=0.5\textwidth]{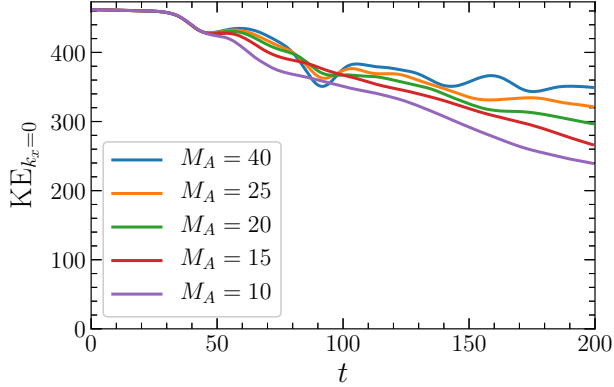}
		\end{center}
		\caption{Kinetic energy over time in the mean flow, $\mathrm{KE}_{k_x=0}$, for a variety of $M_\mathrm{A}$. Each simulation has $\Rm = 500$ and the same. As field strength increases ($M_\mathrm{A}$ decreases), $\mathrm{KE}_{k_x=0}$ decreases more rapidly, equivalent to a faster layer broadening rate.}\label{fig:meanflow_KE_scan}
	\end{figure}

	The shear layer broadens over time in this system as energy is transferred from the mean flow to fluctuations at $k_x>0$. Previous work has shown that the layer broadens more quickly for stronger magnetic fields \cite{Palotti,Mak}. This is consistent with the results shown in Fig.~\ref{fig:meanflow_KE_scan}, where $\mathrm{KE}_{k_x=0}$ is plotted versus time for simulations with the same initial conditions and $\Rm$, but different $M_\mathrm{A}$. As field strength increases, $\mathrm{KE}_{k_x=0}$ decays more rapidly overall. From Eq.~\eqref{eq:meanflow_evolution}, this implies an overall increase in momentum transport down the gradient in $U$, and hence a broadening of the layer. Close inspection shows brief intervals in time where momentum transport reverses and energy is transferred back to the mean flow, as indicated by transient increases in $\mathrm{KE}_{k_x=0}$, e.g., near $t \approx 55$ and $t \approx 85$. While the overall down-gradient transport increases with field strength over long times, the counter-gradient transport in these phases, as well as the down-gradient transport immediately before them, decreases with increased field strength. This will be explored in greater detail in Sec.~\ref{nonlinear:subsec:ModeExcitationTransport}.
	
	The broadening of the layer has important consequences for the eigenmodes and their impact on transport. 
	While the eigenmodes described in Sec.~\ref{sec:eigenmodes} transfer energy to and from the initial base flow and field, the modes governing energy transfer with the background change as the background flow and field change.
	
	In unstable shear layers, the growth rate of the instability generally scales with the difference in flow velocity on either side of the layer divided by the layer width. The most-unstable wavenumber and the critical wavenumber above which modes are no longer unstable scales with one divided by the layer width. 
	Thus, as the layer broadens in time, the critical wavenumber decreases from its initial value of $k_x=1$, and the linear growth rate of fluctuations about the mean flow decreases (see also Ref.~\cite{Hurst2020}, where the inverse is observed in a system where layer thickness is made to decrease over time). Similarly, the Reynolds and Maxwell stresses corresponding to the linear modes also broaden with the layer. 
	
\begin{figure}
		\includegraphics[width=0.5\textwidth]{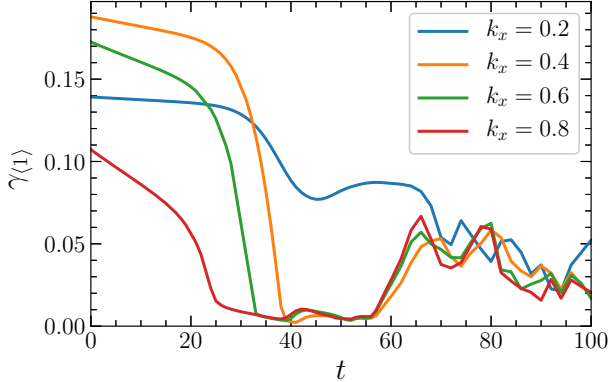}
		\caption[Growth rates over time with quasilinear flattening]{Growth rates for the eigenmodes of $\langle U \rangle_x$ and $\langle B \rangle_x$ taken from a simulation with $M_\mathrm{A} = 40$ and $\Rm = 500$. Different colors correspond to different wavenumbers. For each wavenumber and at each time, only the most unstable growth rate is plotted. Even when perturbations are small and the energy removed from the mean flow appears negligible at early times, the growth rates at higher $k_x$ decline rapidly.}\label{fig:gammas_9_1}
	\end{figure}

	These trends can be shown directly by solving Eqs.~\eqref{eq:linphieq} and \eqref{eq:linpsieq} with $\langle U \rangle_x$ and $\langle B_x \rangle_x$ in place of $U$ and $B_x$, where the mean flow and field are taken from individual timesteps in nonlinear simulations and assumed to be independent of time. Throughout this paper, eigenmodes, complex frequencies, and growth rates obtained in this manner are respectively denoted $\vec{f}_{\langle j \rangle} \equiv (\phi_{\langle j \rangle}(z), \psi_{\langle j \rangle}(z))$, $\omega_{\langle j \rangle}$, and $\gamma_{\langle j \rangle}$ to distinguish them from the eigenmodes of the equilibrium described in Sec.~\ref{sec:eigenmodes}, with $j=1$ and $2$ continuing to denote the most-unstable and conjugate stable modes, respectively, at each wavenumber. Note that Eqs.~\eqref{eq:linphieq} and \eqref{eq:linpsieq} still admit a conjugate stable mode for every unstable mode even when using the mean flow and field, thus $\gamma_{\langle 2 \rangle} = -\gamma_{\langle 1 \rangle}$ still holds. Figure \ref{fig:gammas_9_1} shows how growth rates $\gamma_{\langle j \rangle}$ evolve over time for the most unstable mode at the four initially-unstable wavenumbers for $M_\mathrm{A} = 40$. 
	Even when the mean flow has hardly evolved in the first few timesteps, the growth rates begin to decline noticeably, particularly at the higher wavenumbers. The highest wavenumbers stabilize first, as the critical wavenumber decreases from $k_x=1$. The linear growth regime for $\mathrm{KE}_{k_x=0.4}$ is seen in Fig.~\ref{fig:energies_26_9_1} to end at about the same time that $\gamma_{\langle 1 \rangle} (k_x=0.4)$ approaches zero, suggesting that quasilinear flattening is a dominant saturation mechanism for this mode. The same is true for $k_x=0.6$ and $0.8$ (not shown in Fig.~\ref{fig:energies_26_9_1}). The structure in $z$ of these modes and their corresponding Reynolds and Maxwell stresses are largely the same as $\vec{f}_1$ and $\vec{f}_2$, except that they broaden with the shear layer.
	
	In freely-evolving shear layers, the mean flow has been shown to depart from a simple, broadened $\tanh$ profile in both MHD simulations \cite{Palotti,Mak} and hydrodynamic experiments (see Eq.~(5.2) and Fig.~2 in Ref.~\cite{GKW}), and Ref.~\cite{Wu} (see their Table 1) has shown that even minor departures from a broadened $\tanh$ profile cause significant changes to eigenmode structures and the critical wavenumber. In solving for $\vec{f}_{\langle j \rangle}$ and $\omega_{\langle j \rangle}$, minor features in $\langle U \rangle_x$ and $\langle B_x \rangle_x$ can have a significant impact in the full set of linear modes, including introducing new unstable and conjugate stable modes with finite real frequency. These modes are localized to regions other than $z=0$, including to new inflection points in $\langle U \rangle_x$. At times, changes in $\langle B_x \rangle$ separately introduce new modes as well, consistent with the understanding that nonuniform fields can destabilize shear flows in the absence of inflection points \cite{Tatsuno}. 
	These new modes can form in addition to the existing unstable KH mode, can replace the original conjugate pair of modes with two conjugate pairs of finite-frequency modes, or can emerge after the KH modes have already stabilized at that wavenumber, such as the $k_x \geq 0.4$ unstable modes that emerge around $t \approx 60$ in Fig.~\ref{fig:gammas_9_1}. Each mode's complex frequency is well within the bounds of the modified semicircle theorem derived in Ref.~\cite{Hughes}, but these modes often bear little resemblance to the modes described in Sec.~\ref{sec:eigenmodes}. 
	A detailed investigation of these modes, such as their scaling with different system features, their effects on transport, or their use in reduced models, is beyond the scope of this work. We note the existence of these modes, however, to point out that while the eigenmodes $\vec{f}_{\langle j \rangle}$ of the broadened system correspond more directly to energy transfer to/from the mean than the eigenmodes $\vec{f}_j$ of the equilibrium, analyses in terms of these modes are often unwieldy because of their complexity. Furthermore, as the proceeding subsection will show, the modes of the equilibrium lend themselves to analyses later into the simulation than the modes of the broadened flow.
	
	\subsection{Large-scale eigenmode excitation and momentum transport}\label{nonlinear:subsec:ModeExcitationTransport}
	
	At every time $t$ and wavenumber $k_x$, the Fourier-transformed system state $\hat{f}(k_x,z,t) \equiv (\hat{\phi}(k_x,z,t), \hat{\psi}(k_x,z,t))$ can be expressed as 
	\begin{equation}\label{eq:betadef}
	\hat{f}(k_x,z,t) = \sum_j \beta_j(k_x,t) \vec{f}_j(k_x, z),
	\end{equation}
	provided the eigenmodes of the equilibrium, $\{\vec{f}_j(k_x,z)\}$, form a complete basis. The complex-valued $\beta_j(k_x,t)$ is the amplitude of mode $\vec{f}_j(k_x,z)$ at time $t$ and can be understood as the coefficient of the state vector $\hat{f}$ expressed in this basis. The eigenmodes of the horizontally-averaged system, $\{\vec{f}_{\langle j \rangle}(k_x,z)\}$, can also be used as a basis, and the amplitudes of these modes will be denoted as $\beta_{\langle j \rangle}(k_x,t)$ here. 
	Note that both bases are complete, and the corresponding mode amplitudes are uniquely defined and independent of choice of inner product \cite{FraserThesis}. 
	The procedure for calculating the mode amplitudes $\beta_j$ is essentially identical to that employed in Refs.~\cite{HatchLeft,TerryLeft,Fraser2018}, except that the Laplacian on the left-hand side of Eq.~\eqref{eq:NLphipert} must be taken into account. The same methods are also employed in Ref.~\cite{Burns}.

	\begin{figure}
		\includegraphics[width=1.0\textwidth]{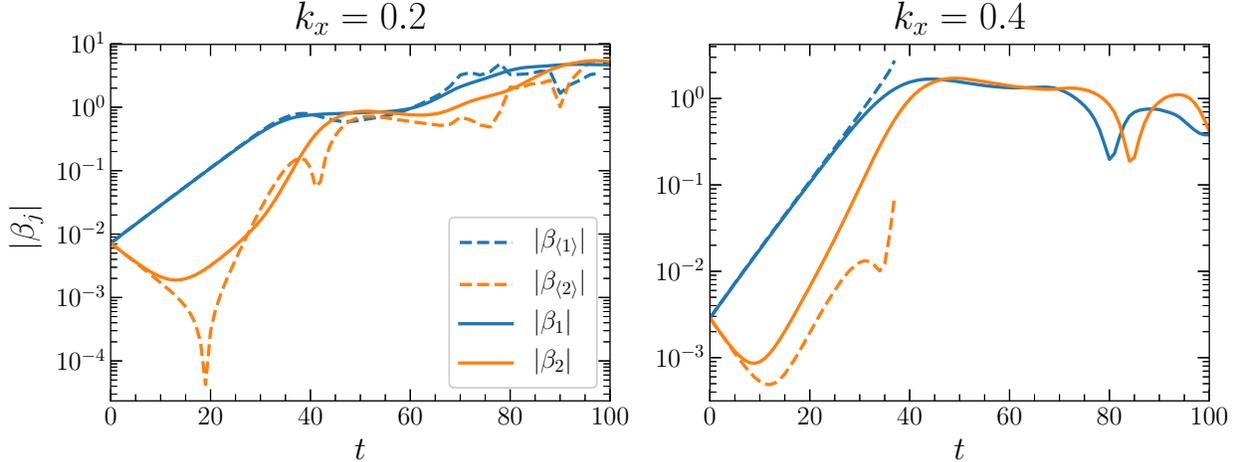}
		\caption[Mode amplitudes, MHD]{Mode amplitudes of the most unstable (blue) and conjugate stable (orange) modes for a simulation with $M_A = 40$ and $\Rm = 500$. Dashed lines are mode amplitudes using a basis of eigenmodes of perturbations about the mean flow and field, solid lines are using eigenmodes of perturbations about the initial equilibrium. For $k_x=0.4$, dashed lines terminate when that wavenumber first stabilizes.
		}\label{fig:betas_9_1}
	\end{figure}
	
	Figure \ref{fig:betas_9_1} shows mode amplitudes over time for the most unstable and conjugate stable mode for the four initially-unstable wavenumbers for the same simulation as in Fig.~\ref{fig:gammas_9_1}, calculated using both sets of modes. At each time, $\beta_{\langle j \rangle}$ is obtained by expanding $\hat{f}$ in terms of the eigenmodes $\{\vec{f}_{\langle j \rangle}\}$ of the instantaneous mean flow and field. Leading into saturation, both sets of amplitudes, $\{\beta_j\}$ and $\{\beta_{\langle j \rangle}\}$, evolve as expected for systems where unstable modes nonlinearly drive stable modes \cite{Terry2006,Makwana,Fraser2017,Fraser2018}: the stable modes decay linearly before being nonlinearly driven while unstable modes are still growing linearly. Comparing the $|\beta_j|$ and $|\beta_{\langle j \rangle}|$ curves demonstrates some of the differences between the two eigenmode bases. 
	Two notable features indicate that $\vec{f}_{j}$ fails to capture the dynamics as precisely as the more relevant $\vec{f}_{\langle j \rangle}$. 
	
	First, there are periods where $|\beta_{2}|$ grows at the same rate as $|\beta_1|$, when it is expected to and eventually does grow faster. These can be understood as follows: suppose the true stable mode amplitude in some system evolves as $\beta_2(t) = \beta_2(0)e^{- |\gamma| t} + \beta_1^2(t)$, and the unstable mode amplitude as $\beta_1(t) = \beta_1(0)e^{|\gamma| t}$ (see, for example, Eqs.~(22-25) in Ref.~\cite{Terry2006}). If a mode that differs slightly from $\vec{f}_2$ is used to calculate $\beta_2(t)$ from simulation data, then an error of the form $\epsilon \beta_1$ will almost always be introduced. This will cause the apparent stable mode amplitude to briefly evolve as $\beta_2(t) \sim \epsilon \beta_1(t)$ before the $\beta_2(t) \sim \beta^2_1$ term becomes dominant, provided that $|\epsilon| \gtrsim |\beta_1(0)|$. For even small differences in $\vec{f}_2$ (measured by some inner product or its corresponding norm), $\epsilon$ can still be quite large provided the eigenmodes are nonorthogonal (under this choice of inner product). 
	Hence, while the differences between the two sets of modes are initially extremely small, they can cause the parametric driving of stable modes by unstable modes to be overlooked in these analyses.
	
	A second subtle effect that is overlooked by the equilibrium modes can be seen when $|\beta_{\langle 2 \rangle}|$ decreases briefly and dramatically before quickly increasing again. This can be attributed to a difference in the complex phases of different interactions (either with the background flow, or with other modes) that determine $\partial\beta_{\langle 2 \rangle}/\partial t$ \cite{Terry2006,Makwana,Fraser2017}: when one interaction overtakes another and becomes dominant, if the two have approximately opposite complex phases, then $\beta_2(t)$ briefly passes through or near $0$ in the complex plane before growing in amplitude.

	\begin{figure}
		\includegraphics[width=0.75\textwidth]{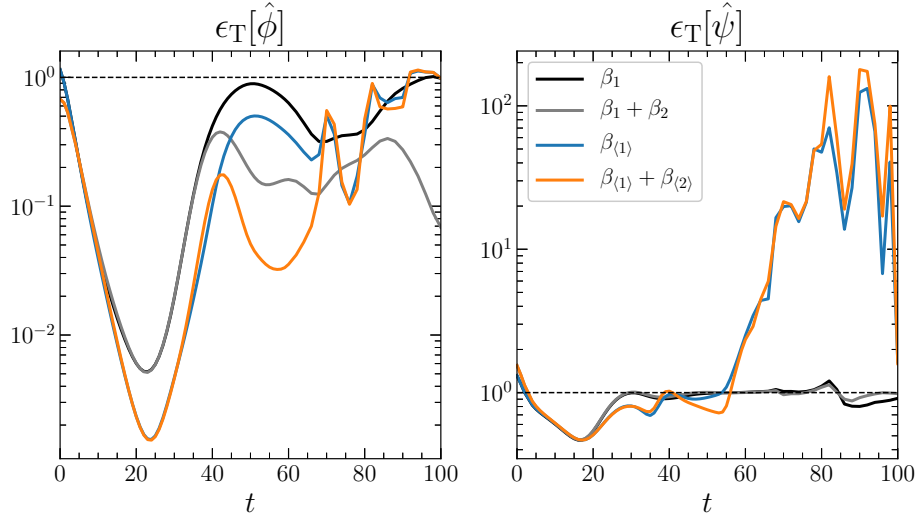}
		\caption[Truncated mode decomposition, $M_\mathrm{A}=40$, $k_x=0.2$]{Error over time for approximations of the flow (left) and field (right) at $k_x=0.2$ using a truncated eigenmode expansion for a simulation with $M_\mathrm{A} = 40$ and $\Rm = 500$. Black and gray curves use $\beta_{j}$ and $\vec{f}_j$, while other curves use $\beta_{\langle j \rangle}$ and $\vec{f}_{\langle j \rangle}$. The horizontal dashed line indicates an error of $1$. For a given approximation, $\epsilon_\mathrm{T} \gtrsim 1$ implies that the difference between the exact state and the approximation is at least as large, energetically speaking, as the state itself, and so the approximation is unreliable. The importance of stable modes can be seen by noting that flow approximations are improved when stable modes are included in both sets of models, with models using $\beta_{\langle j \rangle}$ and $\vec{f}_{\langle j \rangle}$ performing better than those that use $\beta_{j}$ and $\vec{f}_{j}$. Each approximation describes magnetic fluctuations poorly, particularly after the linear regime.}\label{fig:errors_9_1_1}
	\end{figure}
	
	\begin{figure}
		\includegraphics[width=0.75\textwidth]{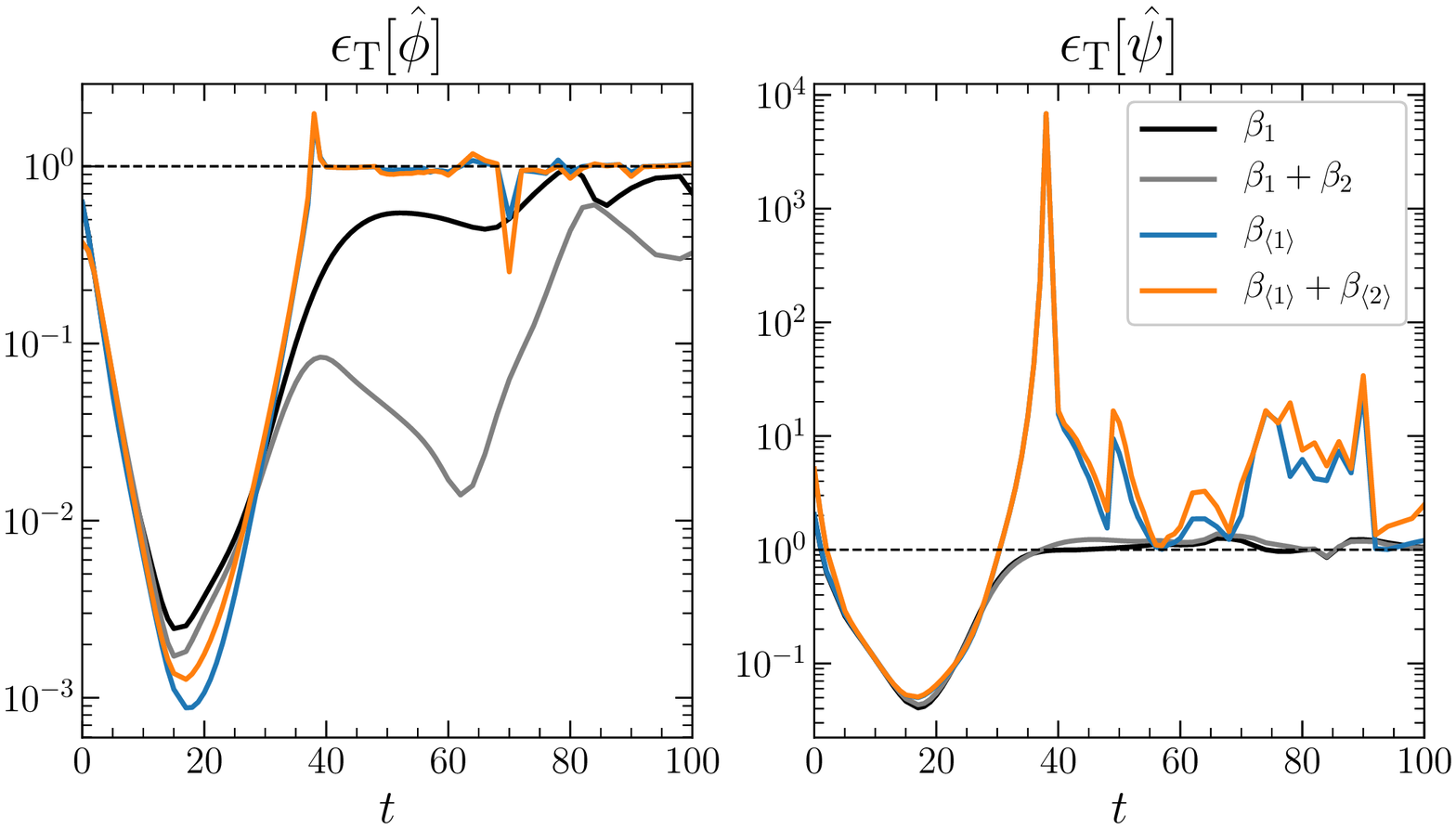}
		\caption[Truncated mode decomposition, $M_\mathrm{A}=40$, $k_x=0.4$]{Identical to Fig.~\ref{fig:errors_9_1_1}, but the $k_x=0.4$ fluctuations are approximated. Unlike the $k_x=0.2$ fluctuations, here the $\vec{f}_{j}$ modes describe $\hat{\phi}$ well for significantly longer than the $\vec{f}_{\langle j \rangle}$ modes. This is to be expected as $\gamma_{\langle j \rangle}$ approaches $0$ much sooner at this $k_x$.}\label{fig:errors_9_1_2}
	\end{figure}
	
	While the mean-flow amplitudes $\beta_{\langle j \rangle}$ capture subtler details leading into saturation than the equilibrium amplitudes $\beta_j$, and the corresponding eigenmodes $\vec{f}_{\langle j \rangle}$ capture the layer broadening, they become unreliable once $\gamma_{\langle 1 \rangle} \to 0$ for a given wavenumber. Furthermore, when additional modes emerge as discussed in Sec.~\ref{nonlinear:subsec:broadening}, identifying broadened modes corresponding to the original stable and unstable modes, and tracking their amplitudes, can become difficult or impossible. 
	Despite the physical effects overlooked by the equilibrium modes, 
	they can still be used to assess stable mode activity and connect it to counter-gradient momentum transport. 
	Their utility in assessing stable mode activity is demonstrated by following similar methods to Ref.~\cite{Fraser2018}: construct two approximations for the $\hat{f}$ by truncating the summation in Eq.~\eqref{eq:betadef} to either include only $j=1$, or $j=1$ and $j=2$. Comparing the accuracy of these two approximations provides a measure of how significant stable modes are in the turbulent state in a way that is normalized to the overall amplitude of the turbulence (whereas $|\beta_2|$ alone is not a normalized measure).
	
\begin{figure}
		\includegraphics[width=0.75\textwidth]{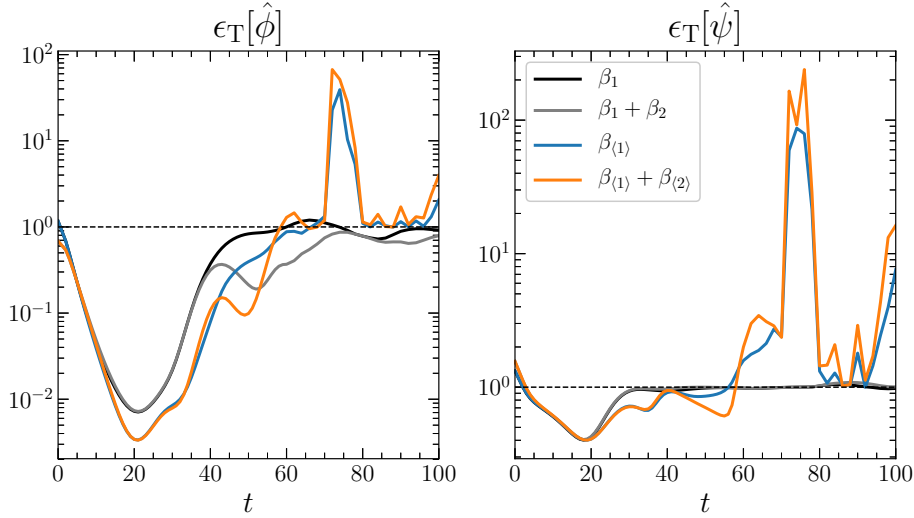}
		\caption[Truncated mode decomposition, $M_\mathrm{A}=7.5$, $k_x=0.2$]{Identical to Fig.~\ref{fig:errors_9_1_1}, but for $M_\mathrm{A} = 7.5$. Here, including stable modes does not improve approximations as noticeably, consistent with their diminished role in the dynamics at lower $M_\mathrm{A}$.}\label{fig:errors_5_1_1}
	\end{figure}

	Figures \ref{fig:errors_9_1_1} and \ref{fig:errors_9_1_2} show, as functions of time, the resulting truncation errors (conceptually similar to, but not to be confused with, the familiar truncation errors in numerical simulations with spectral methods \cite{Boyd}) when fluctuations at $k_x=0.2$ and $0.4$, respectively, are approximated in this manner for a simulation with $M_A = 40$ and $\Rm = 500$. Black and gray curves compare approximations using $\beta_j$ and $\vec{f}_j$, while blue and orange curves compare approximations using $\beta_{\langle j \rangle}$ and $\vec{f}_{\langle j \rangle}$, where $\vec{f}_{\langle 1 \rangle}$ and $\vec{f}_{\langle 2 \rangle}$ are defined as the mode with the largest growth rate and its conjugate. 
	The truncation errors are calculated separately for $\hat{\phi}$ and $\hat{\psi}$ according to 
	\begin{equation}\label{eq:KEerror}
	\epsilon_\mathrm{T}[\hat{\phi}] = \frac{||\hat{\phi}_{\text{exact}} - \hat{\phi}_{\text{approx}}||^2_{\mathrm{KE}}}{||\hat{\phi}_{\text{exact}}||^2_{\mathrm{KE}}}
	\end{equation}
	and $\epsilon_\mathrm{t}$
	\begin{equation}\label{eq:MEerror}
	\epsilon_\mathrm{T}[\hat{\psi}] = \frac{||\hat{\psi}_{\text{exact}} - \hat{\psi}_{\text{approx}}||^2_{\mathrm{ME}}}{||\hat{\psi}_{\text{exact}}||^2_{\mathrm{ME}}},
	\end{equation}
	where $|| \hat{\phi} ||^2_{\mathrm{KE}}$ and $|| \hat{\psi} ||^2_{\mathrm{ME}}$ are the kinetic and magnetic energy of a given $\hat{\phi}$ and $\hat{\psi}$. Unsurprisingly, the $\vec{f}_{\langle j \rangle}$ modes describe flow fluctuations better than the $\vec{f}_j$ ones until quasilinear flattening stabilizes modes at a given wavenumber, at which point such approximations become ill-defined. 
	However, even when quasilinear flattening has set in, flow approximations using eigenmodes of the initial equilibrium retain some fidelity. More importantly, the observation that $\epsilon_\mathrm{T}[\hat{\phi}]$ is reduced when stable modes are included holds true when either set of modes is used. Thus, the equilibrium mode amplitudes $\beta_j$ serve as sufficient indicators of stable mode activity despite the physical effects they overlook. 
	Figure \ref{fig:errors_5_1_1} is identical to Fig.~\ref{fig:errors_9_1_1} but for a simulation with $M_\mathrm{A}=7.5$. Here, including stable modes does not reduce $\epsilon_\mathrm{T}[\hat{\phi}]$ as dramatically as in the $M_\mathrm{A}=40$ case (using either set of modes), suggesting stable modes play less of a role as field strength is increased. Note that in all three figures, each approximation fails to describe the fluctuating field well, particularly after the linear regime. 
	
\begin{figure}
		\includegraphics[width=\textwidth]{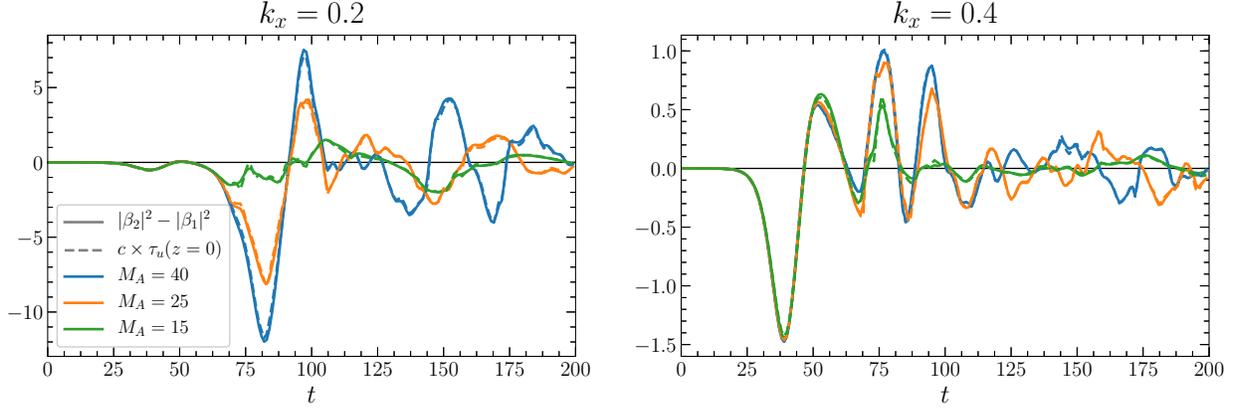}
		\caption{Dashed curves show the contribution to Reynolds stress $\tau_u$ made by fluctuations at $k_x=0.2$ (left) and $0.4$ (right) evaluated at $z=0$ versus time for three of the simulations shown in Fig.~\ref{fig:meanflow_KE_scan}, each with a different value of $M_\mathrm{A}$. Solid curves show the quantity $|\beta_2|^2 - |\beta_1|^2$ over time for the same simulations. The agreement between the two quantities motivates the use of these two modes to characterize the system.}\label{fig:stress_eigenmodes_comparison}
	\end{figure}

	Figures \ref{fig:errors_9_1_1}--\ref{fig:errors_5_1_1} assess the ability of the equilibrium modes $\vec{f}_j$ to describe the state $\hat{f}$ over the entire domain in $z$, demonstrating that they do assess stable mode activity despite the evolving mean flow and field. 
	Focusing on the center of the shear layer, $z=0$, demonstrates that they describe fluctuations there particularly well, and ties stable mode activity to counter-gradient momentum transport. This is shown in Fig.~\ref{fig:stress_eigenmodes_comparison}, where dashed curves show the Reynolds stress at the center of the layer due to fluctuations at different wavenumbers, $\tau_u(k_x,z=0)$, over time for three different $M_\mathrm{A}$. The quantity $|\beta_2|^2 - |\beta_1|^2$ (cf.~Eq.~(23) in Ref.~\cite{Fraser2017} and Eq.~(15) in Ref.~\cite{Fraser2018}) is plotted for comparison. In each case, the two curves essentially overlap, and are nearly identical up to a constant of proportionality that depends on $k_x$ but not $M_\mathrm{A}$. The equilibrium modes $\vec{f}_j$ capture the Reynolds stress at the center of the layer well, and doing so requires only the stable and unstable mode at each wavenumber. Furthermore, whenever stable mode amplitudes exceed unstable mode amplitudes at a given wavenumber, the mid-layer Reynolds stress at that wavenumber changes sign and the stable mode drives counter-gradient momentum transport. Conversely, whenever the mid-layer Reynolds stress changes sign, stable modes exceed unstable modes in amplitude. 
	
\begin{figure}
		\includegraphics[width=0.5\textwidth]{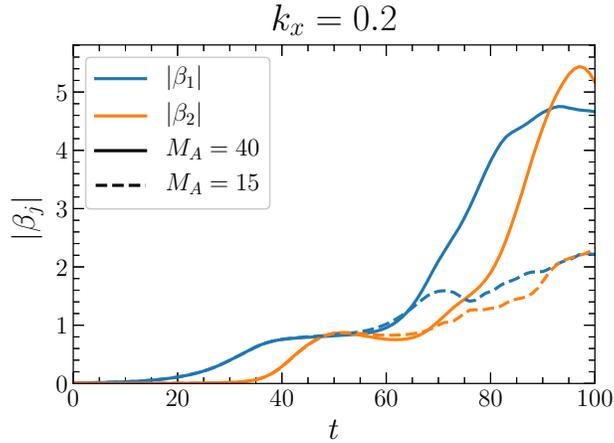}
		\caption{Mode amplitudes $|\beta_2|$ (orange) and $|\beta_1|$ (blue) at $k_x=0.2$ (left) and $0.4$ (right) for $M_\mathrm{A}=40$ (solid), $25$ (dashed), and $15$ (dotted lines). As $M_\mathrm{A}$ decreases, both mode amplitudes are reduced while also trending closer towards equipartition.}\label{fig:beta0s_MAscan}
	\end{figure}

	From Fig.~\ref{fig:stress_eigenmodes_comparison} alone, it is unclear whether the general decrease in $||\beta_2|^2 - |\beta_1|^2|$ with decreasing $M_\mathrm{A}$ is due to a decrease in both mode amplitudes, or whether they are instead trending towards equipartition. The first case might reflect a decrease in transport due to an overall decrease in fluctuation amplitudes, while the second reflects a decrease in transport that is independent of fluctuation amplitudes. Figure \ref{fig:beta0s_MAscan} compares $|\beta_1|$ and $|\beta_2|$ at $k_x=0.2$ for $M_\mathrm{A} = 40$ and $M_\mathrm{A} = 15$, showing that while stable mode amplitudes broadly decrease with increasing field strength, unstable mode amplitudes do as well. As field strength increases the two modes tend towards equipartition. Thus, the reduction in large-scale Reynolds stresses with decreasing $M_\mathrm{A}$ in Fig.~\ref{fig:stress_eigenmodes_comparison} is not purely a result of decreased fluctuation amplitudes. With stronger magnetic fields, the large-scale eddies arrange themselves in a manner that reduces their associated Reynolds stress.
	
\begin{figure}
		\includegraphics[width=0.75\textwidth]{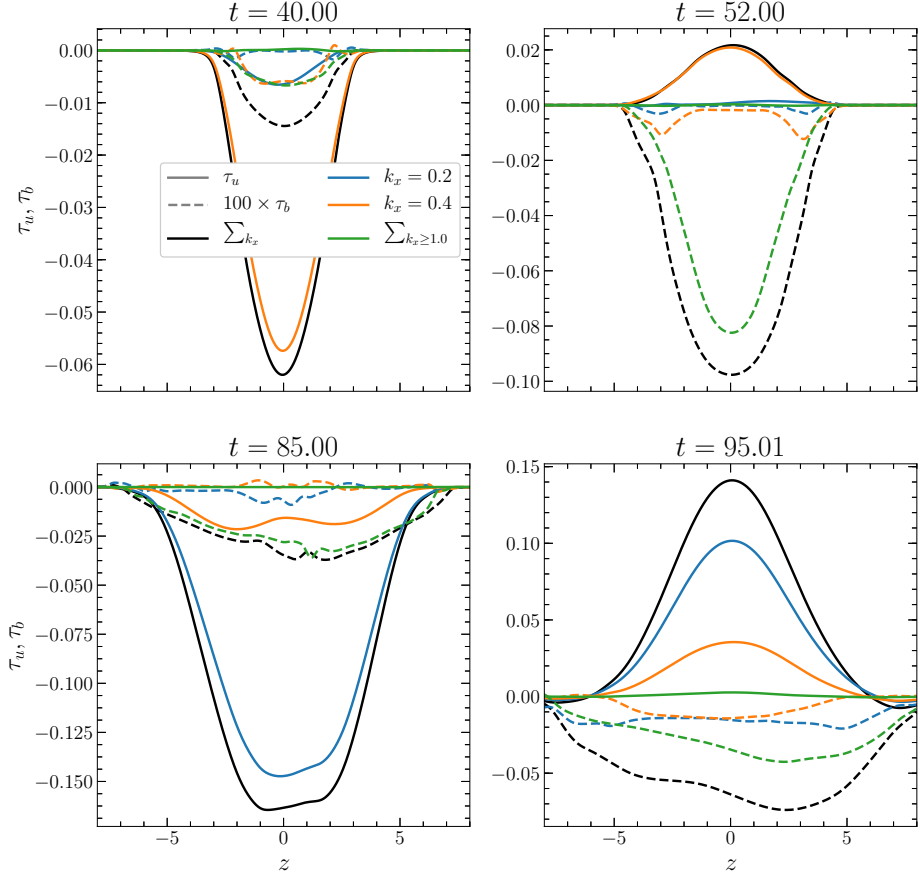}
		\caption{Different contributions to momentum transport at four different times for the same simulation as in Fig.~\ref{fig:contours_26_1}. Solid lines show Reynolds stresses $\tau_u$ and dashed lines show Maxwell stresses $\tau_b$ (rescaled by a factor of $100$ for improved visibility). Black lines are the total Reynolds and Maxwell stresses, while different colors indicate contributions from different $k_x$. While $\tau_b$ is generally dominated by larger $k_x$ and is always negative, $\tau_u$ is dominated by smaller $k_x$ and changes sign over time.}\label{fig:stresses_26_1}
	\end{figure}

	We have established that stable modes are directly responsible for reversals in the mid-layer Reynolds stress of low-$k_x$ fluctuations. To demonstrate that $\tau_u$ is the dominant contributor to the total momentum transport at high $M_\mathrm{A}$, and solely responsible for the local minima in $\mathrm{KE}_{k_x=0}$ identified in Sec.~\ref{nonlinear:subsec:broadening}, we compare $\tau_u$ and $\tau_b$. This is done at four different times for a simulation with $M_\mathrm{A}=60$ and $\mathrm{Rm}=250$ in Fig.~\ref{fig:stresses_26_1}. The four times shown are shortly before and after the first two local minima in $\mathrm{KE}_{k_x=0}$. 
	At each time, the transport is separated into the Reynolds stress $\tau_u$ and Maxwell stress $\tau_b$, which are themselves separated into contributions from different $k_x$. 
	At all times, $|\tau_u|$ exceeds $|\tau_b|$ by over an order of magnitude. The Reynolds stress is dominated first by $k_x = 0.4$ and then by $k_x=0.2$, the same scales that contain the majority of the kinetic energy in $k_x > 0$ fluctuations. Thus, not only are the local minima in $\mathrm{KE}_{k_x=0}$ due to reversals in $\tau_u$ and not $\tau_b$, they are specifically due to $\tau_u$ reversals at small wavenumbers, where we have identified stable-mode excitation as the effect that causes these reversals. This definitively ties counter-gradient momentum transport in this system to stable-mode excitation.
	 
	In contrast to the Reynolds stress, the Maxwell stress $\tau_b$ is dominated by scales beyond the initially-unstable range, aside from the very first panel in Fig.~\ref{fig:stresses_26_1}. This is similarly consistent with the distribution in $k_x$ of magnetic energy. 
	While the low-$k_x$ Reynolds stresses resemble the Reynolds stress of the eigenmodes $\tau_u[\phi_j]$ -- albeit broadened, consistent with the broadening of the flow profile and eigenmodes discussed in Sec.~\ref{nonlinear:subsec:broadening} -- the low-$k_x$ Maxwell stresses bear less of a resemblance to $\tau_b[\psi_j]$, or to $\tau_b[\psi_{\langle j \rangle}]$, particularly as time goes on. Furthermore, $\tau_b$ was never observed to change sign for the simulations considered here. When counter-gradient momentum transport occurs, it is always due to $\tau_u$ changing sign. This is consistent with the truncation error for the field, $\epsilon_\mathrm{T}[\hat{\psi}]$, generally exceeding the truncation error for the flow, $\epsilon_\mathrm{T}[\hat{\phi}]$: if stable and unstable modes alone captured the Maxwell stress as well as they capture the Reynolds stress, then $\tau_b$ at low $k_x$ would reverse sign whenever $\tau_u$ does. 
	Instead, $\tau_b$ consistently transports momentum down the gradient. This $\tau_b$-driven transport is not captured by a simple Fick's law, though more sophisticated models, such as a magnetic eddy viscosity model \cite{Parker}, which applies to the same parameter regime studied here, may perform better.
	
	\begin{figure}
		\includegraphics[width=\textwidth]{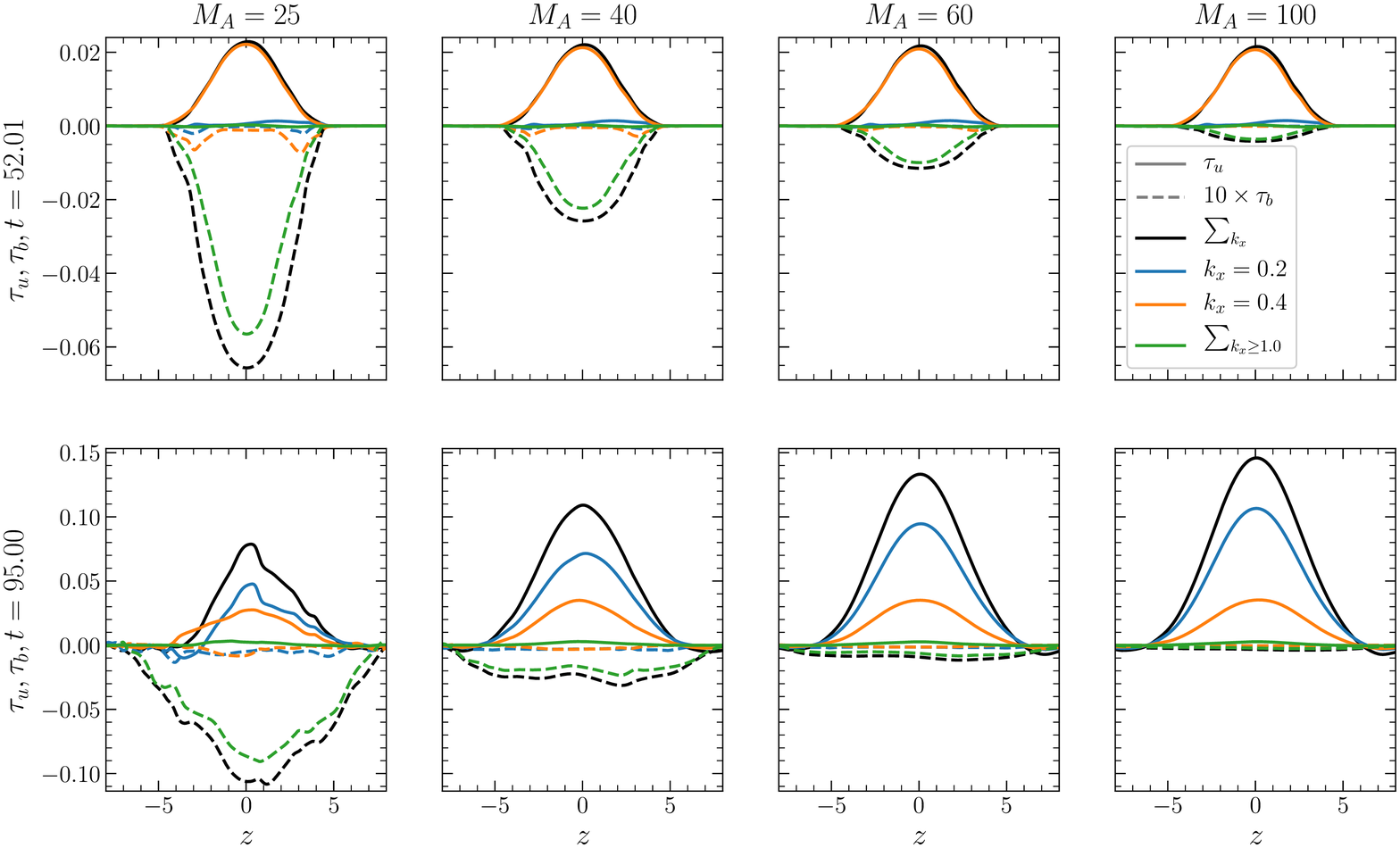}
		\caption[Momentum transport, $M_\mathrm{A}$ scan]{The same breakdown of stresses as in Fig.~\ref{fig:stresses_26_1}, but for four simulations with $\Rm = 500$ and different $M_\mathrm{A}$. The two rows correspond to the first two instances of counter-gradient momentum transport (see, e.g., Fig.~\ref{fig:meanflow_KE_scan}). Note that $\tau_b$ is rescaled by a factor of $10$. As $M_\mathrm{A}$ decreases, $\tau_b$ becomes more dominant, reducing net counter-gradient transport. At earlier times $|\tau_u|$ varies little with $M_\mathrm{A}$, but at later times it decreases with stronger fields.}\label{fig:stresses_scan_highMA_reversed}
	\end{figure}
	
	While counter-gradient momentum transport has been firmly tied to stable mode activity in this paper, the weakening and eventual suppression of counter-gradient transport events with increased field strength is only partly due to magnetic fields reducing stable mode activity. This consistent, down-gradient $\tau_b$ becomes sronger for stronger magnetic fields, and can reduce or even cancel the effect of any counter-gradient $\tau_u$. 
	This is seen in Fig.~\ref{fig:stresses_scan_highMA_reversed}, where $\tau_u$ and $\tau_b$ are shown for different $M_\mathrm{A}$, focusing on the first two counter-gradient transport phases. As field strength increases, $|\tau_b|$ increases. This trend is seen for both a stronger initial field from a decreased $M_\mathrm{A}$ or with more field amplification from an increased $\Rm$ (not shown, however trends are qualitatively similar, but the high-$k_x$ contributions become progressively more dominant at higher $\mathrm{Rm}$). 
	This reduces the net counter-gradient transport during these phases and increases down-gradient transport at other times. 
	In the first phase of counter-gradient transport, $\tau_u$ remains mostly unchanged with $M_\mathrm{A}$, as the relative amplitudes between stable and unstable modes does not significantly change with $M_\mathrm{A}$ at this time (see Fig.~\ref{fig:stress_eigenmodes_comparison}). So while stable mode activity does become less prominent at this time as field strength increases, it is not due to a clear suppression of stable modes relative to unstable ones (cf.~Ref.~\cite{Fraser2018}) and a corresponding decrease in counter-gradient $\tau_u$. 
	Instead, the influence of stable modes has become less prominent due to an increase in energy transfer to small scales leading to increased $|\tau_b(k_x)|$ for large $k_x$ -- that is, the stable modes play a less prominent role only because small-scale magnetic fluctuations are playing more of a role. 
	For $\Rm = 500$, this holds true for all simulations with $M_\mathrm{A} \gtrsim 5$. For lower $M_\mathrm{A}$, the counter-gradient $\tau_u(k_x=0.4)$ at this time does become partially reduced and then fully removed as $M_\mathrm{A}$ decreases, and $\tau_b$ remains down-gradient but becomes dominated by larger scales. 
	In the second phase of counter-gradient transport, the dominant contributor to the Reynolds stress, $k_x=0.2$, becomes noticeably weaker with stronger fields. 
	This trend in $\tau_u$ amplitudes is similar to what happens during the down-gradient transport phases at $t=40$ and $85$: $|\tau_u|$ becomes weaker as field strength increases at later times, but does not change with $M_\mathrm{A}$ at earlier times. 
	
	From these results, the overall increase in down-gradient transport and decrease in $\mathrm{KE}_{k_x=0}$ with field strength can be interpreted as a combination of two factors. First, reduced counter-gradient transport from large-scale flow fluctuations, and second, enhanced down-gradient transport from small-scale field fluctuations. The former is a direct result of decreased stable mode amplitudes. The latter can be understood as an indirect result of stable modes playing a less prominent role in this system. When stable modes activity is significant, energy that would otherwise cascade to small scales is instead returned to the mean flow. As previous work has noted \cite{Makwana2014}, a reduction in stable mode activity allows more energy to reach small scales. The enhanced small-scale fluctuations are further explored in the following subsection.
	
	\subsection{Small-scale fluctuations, dissipation}
	\label{nonlinear:subsec:SmallScales}
	\begin{figure}
		\includegraphics[width=\textwidth]{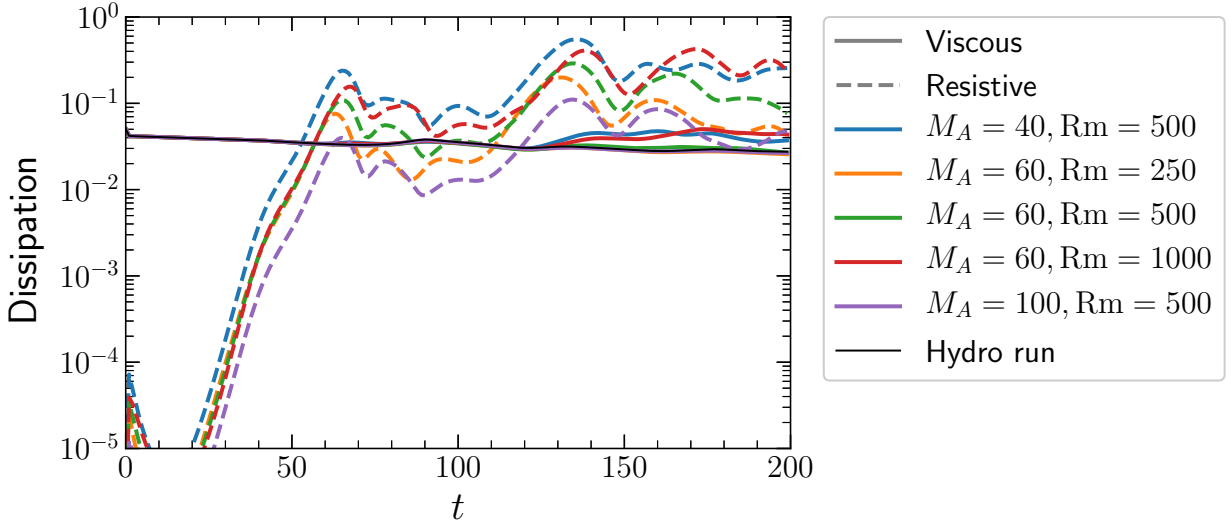}
		\caption[Dissipation timetraces, MHD]{Viscous (solid) and resistive (dashed) dissipation over time for simulations with the same initial conditions as those in Figs.~\ref{fig:hydroNLcontours} and \ref{fig:contours_26_1}, with different colors corresponding to different $M_\mathrm{A}$ and $\Rm$, and black corresponding to the hydrodynamic case. Even a weak magnetic field significantly increases total dissipation relative to the hydrodynamic case. Resistive dissipation increases with $\Rm$, and peaks at a time that increases with $\Rm$. Viscous dissipation is similar to the hydrodynamic case except when the amplified field increases beyond a threshold that depends on both $M_\mathrm{A}$ and $\Rm$.}\label{fig:dissipation}
	\end{figure}

	The enhanced generation of small-scale fluctuations as magnetic field strength increases has been noted in terms of the distribution in $k_x$ of the Maxwell stress (Fig.~\ref{fig:stresses_scan_highMA_reversed}) and of energy (Fig.~\ref{fig:energies_26_9_1}), particularly magnetic energy. When stable mode activity becomes less prominent, the transfer of energy to small, dissipative scales is enhanced. This increases the energy dissipation rate.
	
	Figure \ref{fig:energies_26_9_1} shows that the total energy in the system decreases more rapidly in the MHD simulations than in the hydrodynamic one, and that it decreases more rapidly as field strength increases. The enhanced transfer of energy to small scales coincides with an overall increase in energy dissipation. While the magnetic field does increase the kinetic energy at small scales, the increased dissipation is almost entirely due to the addition of resistive dissipation. This is seen in Fig.~\ref{fig:dissipation}, where viscous and resistive dissipation are plotted for a variety of simulations, each with the same initial condition but different $M_\mathrm{A}$ and $\Rm$. For simulations where the amplified field is not sufficiently strong, the viscous dissipation remains essentially identical to that of the hydrodynamic case. This threshold depends on both $M_\mathrm{A}$ and $\Rm$. For $M_\mathrm{A}=60$, viscosity does not increase until $\Rm$ reaches $1000$, while for $M_\mathrm{A}=40$ the $\Rm = 500$ simulation does exhibit enhanced viscosity while the $\Rm = 250$ case (not shown) does not. The precise threshold for this change in behavior, and a detailed investigation of its cause, is beyond the scope of this paper. 
	Resistive dissipation, on the other hand, matches or even exceeds viscous dissipation, consistent with magnetic energy exceeding kinetic energy at small scales. As expected from Ref.~\cite{ZweibelStagnation} (see Eq.~(6) there), the local maxima in resistive dissipation are greater and occur later with increasing $\Rm$, as the current sheets formed by the high-strain-rate flow take longer to reach small enough scales that resistivity overtakes flux advection. This trend is mostly robust to changes in initial conditions. However, for initial conditions where the merging of the two initial vortices occurs rapidly, the first peak in resistive dissipation may be determined by the time when the merging process disrupts the first current sheet along the braid between the two vortices (e.g.~the current sheet seen at $t=40$ in Fig.~\ref{fig:contours_26_1}), rather than being determined by resistive effects disrupting the sheet \cite{ZweibelStagnation}. For these reasons, care must be taken when studying dissipative processes in KH-unstable systems in MHD to consider appropriate initial conditions, box sizes that permit mergers if relevant, and explicit viscosity \cite{Lecoanet} \textit{and} resistivity \cite{Palotti} wherever possible. 
	While $M_\mathrm{A}$ has no clear effect on the time of these local maxima in high-$M_\mathrm{A}$ simulations, it does affect the overall level of resistive dissipation, with stronger fields (lower $M_\mathrm{A}$) yielding greater resistive dissipation. This can be understood as a lower $M_\mathrm{A}$ increasing the amplitude of the magnetic fluctuations that reach small scales, thus increasing the amplitude of the small-scale, dissipative currents.
	
	\section{Conclusions}
	\label{sec:conclusions}
	
	In turbulence driven by unstable, freely-evolving shear layers in MHD, the addition of a magnetic field has been observed to increase the layer broadening rate due to enhanced turbulent momentum transport\cite{Palotti,Mak}, and to increase energy transfer to small scales, despite stabilizing the driving instability \cite{Chandrasekhar}. This work has investigated the role of large-scale, dissipationless stable modes in this system to identify whether these trends are caused by a reduction in stable mode activity. These modes transfer energy from large-scale fluctuations back to the driving momentum gradient, shrinking the layer width, removing energy from fluctuations, and impeding the cascade to small scales \cite{Makwana2014,Hatch2013}. The results presented here show that the enhanced transfer to small scales and increased broadening rate with stronger magnetic fields do coincide with reduced stable mode activity. 
	Furthermore, in allowing the shear layer to evolve freely without additional forcing terms, this presents the first investigation into the role of stable modes in a system where the source of instability is not fixed\cite{Fraser2017,HatchLeft,Terry2006} or quasi-stationary\cite{Fraser2018}.
	
	Allowing the layer to broaden introduces quasilinear flattening as a distinct saturation mechanism and yields two distinct sets of eigenmodes: the modes obtained by linearizing about the equilibrium flow $U=\tanh(z)$ and field $B_x=1$, denoted by $\vec{f}_j$, and those obtained by linearizing about the instantaneous mean flow $\langle U \rangle_x$ and field $\langle B_x \rangle_x$ at each timestep, denoted by $\vec{f}_{\langle j \rangle}$. The $\vec{f}_{\langle j \rangle}$ modes correspond more directly to energy transfer to and from the mean flow, and they reflect the broadening of fluctuations as the layer broadens. However, they quickly vanish or become ill-defined as the layer evolves and develops small-scale features in $z$. On the other hand, while the $\vec{f}_j$ modes do not correspond to the evolved system as directly as the $f_{\langle j \rangle}$ modes, they remain useful for assessing stable mode activity, and they capture the mid-layer Reynolds stress almost exactly when using only two modes at each $k_x$. 
	Thus, the same eigenmode decompositions used in systems with a fixed \cite{HatchLeft,TerryLeft} or otherwise quasi-stationary \cite{Fraser2018} unstable profile driving the turbulence are also effective in this system, despite the evolution of the mean flow that drives instability.
	 
	These eigenmode decompositions were used to track the amplitudes of stable and unstable modes through the evolving turbulence. In simulations with stronger magnetic fields, stable mode activity becomes less prominent, and the layer broadening rate increases for two reasons. 
	First, stable mode amplitudes are reduced relative to unstable ones in simulations with stronger magnetic fields. This reduces the counter-gradient momentum transport these stable modes drive via turbulent Reynolds stresses. Second, the down-gradient Maxwell stress due to small-scale magnetic fluctuations becomes stronger with stronger fields. This can be understood as a result of stable-mode excitation playing less of a role in instability saturation relative to energy transfer to small scales. When stable modes play less of a role, the energy they would remove from fluctuations instead makes its way to small scales \cite{Makwana2014,Hatch2013}. While the Maxwell stress produced by these small-scale fluctuations does not fit a simple Fick's Law in terms of the mean flow gradient, it is possible that more sophisticated models, such as a magnetic eddy viscosity \cite{Parker}, might capture this effect.
	
	These enhanced small-scale fluctuations, which are generally magnetic fluctuations, can significantly enhance the energy dissipation rate even for very weak magnetic fields (more than doubling the dissipation rate relative to the hydrodynamic case for $M_A = 100$), with dissipation increasing as $\Rm$ increases or $M_A$ decreases. 
	This increased dissipation is almost entirely due to resistive dissipation, although viscous dissipation increases with field strength as well, provided the field is stronger than some threshold that depends on $\Rm$ and whose precise value for different $\Rm$ is beyond the scope of this paper. Thus, even in the presence of a magnetic field much too weak to significantly affect the linear instability, shear layers in non-ideal MHD seem to dissipate energy much faster than the hydrodynamic counterpart, with the reduction of stable mode activity providing an underlying explanation.
	
	This work motivates two related possible future directions of inquiry. First, while the Reynolds stresses in this system can be reasonably modeled using stable and unstable modes alone, the Maxwell stresses cannot. Not only do these modes fail to describe large-scale field fluctuations well, but the Maxwell stress is primarily driven by small-scale fluctuations, while the modes considered in this paper exist at large scales. Recent work \cite{Parker} involving a quasi-stationary system of driven, shear-flow turbulence has suggested these stresses can be described by a magnetic eddy viscosity model. It is plausible that for the same system considered here, with the addition of a shear forcing term to permit a quasi-stationary state, the momentum transport could be well-described by a combination of stable and unstable modes for the Reynolds stress and a magnetic eddy viscosity model for the Maxwell stress. Furthermore, a quasi-stationary system enables a thorough investigation of the dominant nonlinear interactions in the saturated state, potentially informing saturation theories for predicting eigenmode amplitudes without first performing direct numerical simulations (similar to the saturation theories \cite{Terry2018} that were informed by studies of dominant nonlinear interactions \cite{Makwana2014} in the context of fusion plasmas). Provided a reduced model that is informed by stable mode activity, a second direction for this work is the application of such a model to physical systems. For example, if the effect of density stratification is included, such a model might be used to improve predictions of shear-driven transport in stellar evolution codes \cite{Paxton}.

	\acknowledgments{
		The authors thank K.~Burns, D.~Lecoanet, C.~Sovinec, and J.~Parker for insightful discussions, and N.~Hurst for insightful discussions and a thorough reading of the manuscript. The authors also thank the Dedalus developers and the Dedalus user group for assistance with many aspects of installing and running Dedalus. Partial support for this work was provided by the National Science Foundation under Award PHY-1707236, and by the U.S.~Department of Energy, Office of Science, Fusion Energy Sciences, under Award No.~DE-FG02-04ER54742. Computing resources were provided by the National Science Foundation through XSEDE computing resources, allocation Nos.~TG-PHY130027 and TG-PHY180047. This research was performed using the compute resources and assistance of the UW-Madison Center For High Throughput Computing (CHTC) in the Department of Computer Sciences. The CHTC is supported by UW-Madison, the Advanced Computing Initiative, the Wisconsin Alumni Research Foundation, the Wisconsin Institutes for Discovery, and the National Science Foundation, and is an active member of the Open Science Grid, which is supported by the National Science Foundation and the U.S. Department of Energy's Office of Science.}

\bibliography{StableModes_MHD_KH_2020}
\end{document}